\documentclass[usenatbib,psfig]{mn2e}

\usepackage{epsfig}

\begin{document}


\title[Diffuse LINER emission from barred galaxies]{Diffuse LINER-type emission from extended disc regions of barred galaxies}
\author[S. M. Percival \& P. A. James]{S. M. Percival\thanks{E-mail:
S.M.Percival@ljmu.ac.uk} \& P. A. James\\
Astrophysics Research Institute, Liverpool John Moores University, IC2, Liverpool Science Park, 146 Brownlow Hill, Liverpool L3 5RF, UK\\
}

\date{Accepted 2020 May 13. Received 2020 May 13; in original form 2019 August 15}

\pagerange{\pageref{firstpage}--\pageref{lastpage}} \pubyear{2020}

\maketitle

\label{firstpage}

\begin{abstract}
 We present a spectroscopic analysis of the central disc regions of barred spiral galaxies, concentrating on the region that is swept by the bar but not including the bar itself (the `Star Formation Desert' or SFD region). New spectroscopy is presented for 34 galaxies, and the full sample analysed comprises 48 SBa - SBcd galaxies.  These data confirm the full suppression of star formation within the SFD regions of all but the latest type (SBcd) galaxies.  However, diffuse [N{\sc ii}] and H$\alpha$ line emission {\em is} detected in all galaxies.  The ubiquity and homogeneous properties of this emission from SBa - SBc galaxies favour post-Asymptotic Giant Branch (p-AGB) stars as the source of this line excitation, rather than extreme Blue Horizontal Branch stars.  The emission-line ratios strongly exclude any contribution from recent star formation, but are fully consistent with recent population synthesis modelling of p-AGB emission by other authors, and favour excitation dominated by ambient gas of approximately solar abundance, rather than ejecta from the AGB stars themselves.  The line equivalent widths are also larger than those observed in many fully passive (e.g. elliptical) galaxies, which may also be a consequence of a greater ambient gas density in the SFD regions. 
\end{abstract}

\begin{keywords}
Galaxies : active - galaxies : spiral - galaxies : stellar content - galaxies : structure
\end{keywords}

\section{Introduction}

The presence of a bar in the centre of a disc galaxy has profound consequences for the evolution of that galaxy, with effects including the redistribution of gas, triggering, redistribution and quenching of star formation (SF), and the development of long-lived stellar components not seen in unbarred galaxies \citep{frie93,korm04,korm13}. The timescales and interplay of these various gas-phase and stellar processes have yet to be determined in detail, however.  In a series of papers, we have been looking at one specific mode of SF quenching that is intimately associated with strong bars in high-mass spiral galaxies.  This was first seen through the discovery of strong suppression of H$\alpha$ line emission in some of the galaxies studied in the H$\alpha$ Galaxy Survey \citep[H$\alpha$GS; ][]{jame04}, with the suppression being limited to the region of the disc `swept' by a strong bar \citep{jame09}.  The resulting `Star Formation Desert' (SFD) as we termed it has significant global consequences for the galaxies concerned, as the affected area (a radial range extending approximately 1--5\,kpc from the nucleus) is just the region that dominates disc SF in unbarred galaxies. Thus this represents bars at least redistributing, and possibly suppressing, a significant fraction of the total SF in each affected galaxy.

In \cite{jame15} we presented the first spectroscopic study of SFD regions, using long-slit optical spectra of 15 nearby barred galaxies to confirm that SFD regions are indeed devoid of SF activity (although nuclei at the bar centres and rings around the peripheries of bars are frequently host to vigorous SF).  However, we found that SFD regions often contain diffuse low-surface-brightness line emission, with line ratios showing excitation that is clearly inconsistent with that expected from young stars.  The extent of this emission, generally filling the SFD regions, argues against Active Galactic Nuclei (AGN) as the origin of the excitation, and our preferred explanation was instead the hard radiation field expected to result from post-Asymptotic Giant Branch (p-AGB) stars \citep{cald84,phil86,goud99,grav07,sing13,brem13}. For at least one galaxy in our study, this diffuse extended line emission shares fully in the rotational velocity of the disc of its host galaxy, supporting a link with the stellar component rather than an extended halo.

In subsequent papers \citep{jame16,jame18} we analysed absorption-line spectroscopy of SFD regions to determine stellar population ages and metallicities.  However, in the present paper, we return to the emission-line properties of these regions, using new data which extend the sample size from 15 galaxies in \cite{jame15} to 48 galaxies, and improve the spectroscopic coverage of some of the galaxies already observed.  With this improved dataset, we can explore further the ubiquity of the SFD phenomenon in barred galaxies, determine how common is the association of low-level diffuse line emission with these regions, and test the idea that that such emission is powered by p-AGB stars against other possibilities, such as bar-driven shocks or diffuse ionized gas (DIG) excited by leakage of ionizing photons from H{\sc ii} regions \citep{haff09}. Other questions include the mechanism of the SF suppression; is this due to the efficient removal of gas from SFD regions, or is the gas still present but stabilised, e.g. by turbulent velocities generated by the bars?

\subsection{Modelling line emission from evolved stellar populations}

There is an extensive literature on both observational and theoretical studies of stellar-driven line emission that is not driven by star formation.  A central question concerns the stellar evolutionary phase of the stars that dominate this emission, with the main candidates being p-AGB stars and Extreme Blue Horizontal Branch (EHB) stars. Here we briefly introduce some of the main studies that are relevant for the present analysis.

In an early study, \cite{bine94} demonstrated that
p-AGB stars provide sufficient ionising radiation to account for the small but detectable H$\alpha$ luminosities and equivalent widths observed in some early-type galaxies.  \cite{stas08} introduced the concept of `retired' galaxies, i.e. those with no ongoing SF, and used population synthesis modelling to demonstrate that they can have detectable line emission due to p-AGB stars and hot white dwarfs, which may explain a substantial fraction of LINER-type galaxies, without recourse to AGN-driven excitation.
\cite{cidf11} extended this analysis, and applied it to a large sample of galaxies for which spectroscopy of the central regions was available from the Sloan Digital Sky Survey (SDSS).  They applied a diagnostic diagram using H$\alpha$ equivalent width (EW) and [N{\sc ii}]/H$\alpha$ line ratios to the classification of sources into star-forming, AGN, `retired' and completely passive categories, finding a bimodal distribution in H$\alpha$ EW that distinguishes AGN from evolved star (assumed to be p-AGB) ionisation with a borderline at 3\,\AA; completely passive galaxies were found to have equivalent widths $<$ 0.5\,\AA.

\cite{belf16} analysed data for 646 galaxies from the SDSS-IV MANGA survey, showing extended emission with LINER-type ratios in many of these. Again, they argued for p-AGB stars as the ionising source, and concluded that shock-excited galaxies are rare. 
\cite{papa13} and \cite{gome16} presented a study of 32 early-type galaxies from the CALIFA survey, finding two classes of extended emission, both with LINER-type ratios.  The classes are differentiated by their H$\alpha$ EW and radial profiles, with their type i group exhibiting stronger line emission (EW $>$ 0.5\AA) and steep radial gradients in this EW.  This emission was attributed to p-AGB excitation, while the type ii galaxies with low EW and shallower emission gradients were concluded to be dominated by ionizing photon leakage. \cite{hirs17} presented modelling of changes in emission line ratio as a function of cosmological epoch, and found p-AGB excitation to be important at lower redshifts, with AGN becoming more dominant at earlier cosmological epochs.

One study that is particularly relevant to the present paper is that of \cite{byle17,byle19}, who have carried out probably the most comprehensive modelling of the emission-line properties of evolved stellar populations.  Their analysis is based on MIST \citep{choi16}, which in turn builds on the MESA stellar synthesis models \citep{dott16} plus the CLOUDY ionisation code \citep{ferl13}.  \cite{byle19} find p-AGB emission to be ubiquitous in all but the very youngest populations; it is initiated at $\sim$10$^8$ years, and the continues until population ages of at least 10$^{10}$ years, with a slow decline in total ionising flux as a function of population age.  However, as the population itself fades, the predicted H$\alpha$ EW actually rises slowly with age, in the range $\sim$0.2 - 2.5\,\AA.  The use of CLOUDY enables them to predict line ratios from the gas around these p-AGB populations.  If they assume ambient, pre-existing gas they find [N{\sc ii}]/H$\alpha$ line ratios that are clearly higher than those predicted for star-forming regions, but that lie in the intermediate range between star formation and the classic LINER-type regime, as defined by e.g. \cite{kewl06}. However, the models of \cite{byle17,byle19} do predict LINER-type ratios if they use what they call $\alpha$CN abundance ratios, such that the gas is enhanced in $\alpha$ elements, carbon, and nitrogen, which they argue would be expected if the local environment of each p-AGB star is substantially enriched by gas lost by the star during its AGB evolution.

Another possible source of extended line emission is DIG, where the ionizing photons escape from H{\sc ii} regions, producing a low-surface-brightness floor of ionization on scales as large as kiloparsecs \citep{zuri02,zhan17}. \cite{belf16} find that this emission can have LINER-like emission line ratios, and \cite{ferg96} claim that this extended component can comprise up to 50\% of the H$\alpha$ emission, even in star-forming galaxies.  Recent studies have confirmed this diffuse emission using deep Integral Field Unit observations; \cite{denb20} study 41 galaxies with the MUSE instrument on the ESO Very Large Telescope, finding a diffuse H$\alpha$ fraction of $\sim$30\% for one of the fitting methods they use, which they interpret as DIG.  \cite{weil18} also use MUSE in a study of the central regions and an outer tidal arm of `The Antennae', NGC\,4038/4039, again finding a diffuse component which they conclude can be excited by leakage of ionizing photons.

SFD regions provide a unique environment to test models of line emission powered by sources other than young stars, as they involve the central regions of gas-rich galaxies, but with no trace of the star formation which completely dominates such emission in most gas-rich galaxies. 
In this paper, we extend our earlier sample of 15 barred galaxies by adding deep long-slit spectroscopy for 34 systems.  One of these had spectroscopy in our previous paper, and hence the combined dataset comprises 48 galaxies.  Multiple slit position angles for several galaxies enable a comparison of the line emission at additional positions within some of the SFD regions, and investigation of the spectroscopic appearance of bars themselves for comparison with their surrounding SFD regions.

The remainder of this paper is structured as follows. Section 2 contains a description of the galaxy sample, the new spectroscopic observations, and a brief account of the data reduction.  The results of the spectroscopic analysis are given in Section 3, followed by a discussion of the consequences of these observations for our understanding of the SFD phenomenon in Section 4.  Our conclusions are summarised in Section 5.

\section{Observations}

The 34 galaxies for which new spectroscopic data are presented here are listed in Table~\ref{tab:gals_obs}.  The first 6 columns list the following galaxy properties: names, classifications, nuclear properties as listed in the NASA Extragalactic Database, recession velocities, adopted distances, and bar position angles in degrees anticlockwise from a North-South orientation. Columns 7 - 10 give some details of our observations: dates, integration times in seconds, slit position angles and the outer radius in kpc of the region extracted in our SFD spectroscopy.  This sample supplements that presented in \cite{jame15}.  Compared with our earlier study, the new sample is substantially larger, contains galaxies at broadly similar distances \citep[mean 58.5\,Mpc, range 14.1 -- 122.8\,Mpc, cf. a mean of 39.2\,Mpc, range 10.2 -- 95.5\,Mpc for][]{jame15}. The sample somewhat extends the Hubble-type range, in that we now have 6 early-types (SBa/SBab) cf. only one in the earlier study, and 5 late-types, classified SBcd, again cf. one previously.

The new observations presented here are long-slit optical spectra, taken with the Intermediate Dispersion Spectrograph (IDS) on the 2.5-metre Isaac Newton Telescope (INT) at the Roque de Los Muchachos Observatory on La Palma in the Canary Islands.  The observing time was allocated to proposal I/2015A/07 by the UK Panel for the Allocation of 
Telescope Time. Observing dates allocated were 13 - 20/03/2015, but no data taken on last three nights due to poor weather. IDS was used with grating R1200Y, which has a blaze wavelength of 6000\,\AA. The detector used was Red+2, optimised for the longer optical wavelengths of primary interest here.  The slit width used for all galaxy observations was 1.5$^{\prime\prime}$, giving a spectral dispersion corresponding to  0.53\,\AA/pixel and 
wavelength coverage of 1144\,\AA; the selected central wavelength was 6604\,\AA. 
The spatial pixel scale was 0.44$^{\prime\prime}$/pixel.
We note that the slit position angles were set with respect to the bar axis, e.g. usually perpendicular to the bar for SFD observations, and so were {\em not} at the parallactic angle; however, atmospheric refraction effects are negligible for the [N{\sc ii}]/H$\alpha$ line ratios that are used here.

Flux calibration came from observations of spectrophotometric standard stars selected from the list at\\ http://catserver.ing.iac.es/landscape/tn065\-100/workflux.php.

\begin{table*}
 \begin{minipage}{140mm}
  \caption{The galaxy sample and observational details for the new observations presented here. Position angles in italics are for observations along the axis of the bar.}
  \begin{tabular}{llcrrrrrcr}
  \hline
 Name     &  Class       &    Nucl(NED)     &   Vel    & Dist       & Bar     &   Date obs  & Int time        &  Slit PA         & R$_{\rm MAX}$     \\
          &              &                  &  km/s    & (Mpc)      & PA      &             &                 &                  & (kpc)    \\
  \hline
 NGC1924  &   SB(r)bc    &                  &   2558   &    36.0    &    45   &   20150315  &   3$\times$1200        &  315             &  2.3     \\
 NGC2326  &   SB(rs)b    &                  &   5985   &    90.4    &   130   &   20150317  &   3$\times$1200        &  220             &  8.7     \\
 NGC2339  &   SAB(rs)bc  &        AGN       &   2206   &    33.0    &    70   &   20150313  &   3$\times$1200        &  340             &  2.1     \\
 UGC3973  &   SBb        &        Sy1.2     &   6652   &   100.2    &    70   &   20150314  &   3$\times$1200        &  340             &  4.5     \\
 UGC4042  &   (R)SB(r)b  &                  &   8292   &   122.8    &   150   &   20150315  &   3$\times$1200        &   60             &  6.2     \\
 NGC2487  &   SB(r)c     &                  &   4841   &    71.9    &    45   &   20150314  &   3$\times$1200        &  {\it 43},315          &  6.8     \\
 NGC2545  &   (R)SB(r)ab &        LINER     &   3385   &    50.6    &   170   &   20150317  &   3$\times$1200        &  260             &  1.7     \\
 NGC2523  &   SB(r)bc    &                  &   3471   &    55.3    &   120   &   20150313  &   3$\times$1200        &   60             &  7.1     \\
 NGC2604  &   SB(rs)cd   &    WR,H{\sc ii},Sbrst  &   2078   &    32.7    &    50   &   20150316  &   (1)3$\times$1200     &  {\it 230},320         &  2.8     \\
 NGC2746  &   SB(rs)a    &        AGN?      &   7065   &   105.6    &    40   &   20150315  &   3$\times$1200        &  310             &  8.1     \\
 NGC3049  &   SB(rs)ab   &       H{\sc ii},Sbrst  &   1455   &    24.2    &    35   &   20150316  &   3$\times$1200        &  305             &  3.5     \\
 NGC3185  &   (R)SB(r)a  &        Sy2       &   1217   &    22.9    &   120   &   20150317  &   3$\times$800         &  30,285,{\it 300},315  &  3.5     \\
 NGC3346  &   SB(rs)cd   &                  &   1274   &    22.4    &    90   &   20150313  &   3$\times$1200        &    0             &  1.2     \\
 UGC5892  &   SBb        &                  &   8107   &   119.7    &    90   &   20150315  &   3$\times$1200        &  180             &  3.3     \\
 NGC3374  &   SBc        &                  &   7468   &   112.7    &    15   &   20150314  &   3$\times$1200        &  285             &  4.1     \\
 NGC3485  &   SB(r)b:    &        AGN       &   1436   &    25.7    &    40   &   20150313  &   3$\times$1200        &  130             &  1.8     \\
 NGC3507  &   SB(s)b     &        LINER     &    979   &    15.8    &   120   &   20150316  &   3$\times$1200        &   30             &  2.0     \\
 NGC3729  &   SB(r)a pec &                  &   1060   &    20.2    &    20   &   20150315  &   3$\times$1200        &  290             &  1.4     \\
 NGC3963  &   SAB(rs)bc  &                  &   3188   &    51.5    &   130   &   20150314  &   3$\times$1200        &  220             &  4.3     \\
 NGC4123  &   SB(r)c     &    WR,H{\sc ii},Sbrst  &   1327   &    19.3    &   105   &   20150316  &   3$\times$1200        &   15             &  2.5     \\
 NGC4416  &   SB(rs)cd   &      H{\sc ii},Sbrst   &   1390   &    14.1    &     5   &   20150317  &   1$\times$1200        &   95             &  0.6     \\
 NGC4619  &   SB(r)b     &         Sy1      &   6927   &   105.2    &    10   &   20150313  &   3$\times$1200        &  100             &  3.2     \\
 NGC4779  &   SB(rs)bc   &     H{\sc ii},Sbrst    &   2831   &    44.6    &    10   &   20150316  &   3$\times$1200        &  100             &  3.9     \\
 NGC4904  &   SB(s)cd    &   H{\sc ii},WR,Sbrst   &   1180   &    20.8    &   135   &   20150314  &   3(1)$\times$1200     &   45,{\it 315}         &  2.3     \\
 NGC4999  &   SB(r)b     &                  &   5647   &    84.7    &    60   &   20150315  &   3$\times$1200        &  330             &  8.5     \\
 NGC5164  &   SBb        &                  &   7218   &   110.5    &    25   &   20150317  &   3$\times$1200        &  115             &  3.8     \\
 NGC5350  &   SB(r)b     &        Sbrst     &   2321   &    39.3    &   110   &   20150313  &   3$\times$1200        &   20             &  3.2     \\
 NGC5375  &   SB(r)ab    &                  &   2386   &    39.9    &   170   &   20150316  &   3$\times$1200        &  260             &  3.9     \\
 NGC5618  &   SB(rs)c    &                  &   7132   &   106.7    &     5   &   20150317  &   3$\times$1200        &   95             &  3.6     \\
 IC1010   &   SB(r)b pec &                  &   7702   &   115.3    &   170   &   20150315  &   3$\times$1200        &   80             &  6.7     \\
 NGC5735  &   SB(rs)bc   &                  &   3742   &    59.3    &     0   &   20150314  &   (1)3$\times$1200  &    {\it 0},270         &  3.3     \\
 IC1067   &   SB(s)b     &                  &   1577   &    27.5    &   140   &   20150313  &   3$\times$1200        &   45             &  1.7     \\
 NGC5970  &   SB(r)c     &     H{\sc ii},LINER    &   1957   &    28.4    &    80   &   20150315  &   (1,1,1)3$\times$1200 &  65,{\it 80},95,350    &  1.8     \\
 NGC6004  &   SAB(rs)bc  &                  &   3826   &    60.3    &    10   &   20150317  &   3$\times$1200        &  100             &  2.3     \\
\hline
\end{tabular}
\label{tab:gals_obs}
\end{minipage} 
\end{table*}

\begin{table*}
 \begin{minipage}{140mm}
  \caption{The observed regions and extracted spectroscopic parameters for the full sample of 48 galaxies.}
  \begin{tabular}{lrlcccccccccc}
  \hline
  
  Name   & Slit   &   Region &   EW$_{\rm{H\alpha}}$   &   err   & EW$_{\rm{[N{\sc ii}]}}$  &   err   & [N{\sc ii}]/H$\alpha$      &  err      & [N{\sc ii}]/H$\alpha$      &     err   &  Rad range  & Dir  \\
         &  ~PA   &          &          &         &         &         & ~(obs)      &           & ~(cor)      &           &     $^{\prime\prime}$        &       \\
\hline
 NGC 1924  &  315   &   SFD1   &    2.42  &   0.14  &   2.51  &   0.24  &   1.038   &  0.111  &   0.668     &   0.108   &   3.1-6.2   &  NW   \\
 NGC 1924  &  315   &   SFD2   &    3.21  &   0.14  &   2.28  &   0.12  &   0.711   &  0.067  &   0.501     &   0.067   &   4.4-13.2  &  SE   \\
 NGC 1924  &  315   &   H{\sc ii}    &   50.52  &   0.25  &  20.15  &   0.20  &   0.399   &  0.011  &   0.389     &   0.011   &  16.7-22.4  &  SE   \\
 NGC 2326  &  220   &   SFD1   &    ---   &   ---   &   1.43  &   0.29  &    $>$1     &  ---    &   1.068     &   0.228   &   5.7-19.8  &  SW   \\
 NGC 2326  &  220   &   SFD2   &    ---   &   ---   &   ---   &   ---   &    N/A    &  ---    &    N/A      &   ---     &   5.7-19.4  &  NE   \\
 NGC 2326  &  220   &   H{\sc ii}    &    7.84  &   0.47  &   3.27  &   0.26  &   0.417   &  0.100  &   0.356     &   0.096   &  20.7-25.1  &  SW   \\
 NGC 2339  &  340   &   SFD1   &    2.56  &   0.10  &   2.75  &   0.19  &   1.074   &  0.079  &   0.705     &   0.080   &   7.5-13.2  &  S    \\
 NGC 2339  &  340   &   SFD2   &    2.53  &   0.16  &   2.83  &   0.25  &   1.122   &  0.109  &   0.733     &   0.104   &   8.4-11.9  &  N    \\
 NGC 2339  &  340   &   H{\sc ii}    &   25.43  &   0.14  &  10.27  &   0.16  &   0.404   &  0.017  &   0.384     &   0.018   &  18.0-24.2  &  S    \\
 UGC 3973  &  340   &   SFD    &    1.26  &   0.18  &   2.58  &   0.29  &   2.050   &  0.179  &   0.993     &   0.140   &   7.0-9.2   &  SE   \\
 UGC 3973  &  340   &   H{\sc ii}    &    4.21  &   0.24  &   2.57  &   0.31  &   0.610   &  0.135  &   0.463     &   0.132   &  10.1-13.2  &  SE   \\
 UGC 4042  &   60   &   SFD1   &    1.68  &   0.20  &   1.62  &   0.18  &   0.965   &  0.163  &   0.536     &   0.136   &   3.3-10.4  &  NE   \\
 UGC 4042  &   60   &   SFD2   &    1.85  &   0.16  &   1.62  &   0.22  &   0.872   &  0.161  &   0.506     &   0.151   &   3.5-8.4   &  SW   \\
 UGC 4042  &   60   &   H{\sc ii}    &   25.28  &   0.32  &  10.53  &   0.27  &   0.416   &  0.028  &   0.395     &   0.028   &   9.2-11.9  &  SW   \\
 NGC 2487  &  315   &   SFD1   &    1.14  &   0.16  &   2.10  &   0.23  &   1.840   &  0.176  &   0.876     &   0.138   &   6.2-19.4  &  SE   \\
 NGC 2487  &  315   &   SFD2   &    ---   &   ---   &   1.32  &   0.14  &    $>$1     &  ---    &   0.984     &   0.147   &   6.6-14.1  &  NW   \\
 NGC 2487  &  315   &   H{\sc ii}    &   16.03  &   0.22  &   6.31  &   0.25  &   0.394   &  0.042  &   0.363     &   0.043   &  22.4-29.0  &  SE   \\
 NGC 2487  &   43   &   Bar1   &    ---   &   ---   &   1.07  &   0.12  &    $>$1     &  ---    &   0.801     &   0.150   &   6.2-16.7  &  SW   \\
 NGC 2487  &   43   &   Bar2   &    ---   &   ---   &   1.19  &   0.19  &    $>$1     &  ---    &   0.891     &   0.190   &   6.2-17.2  &  NE   \\
 NGC 2487  &   43   &   H{\sc ii}    &   56.11  &   0.39  &  20.21  &   0.26  &   0.360   &  0.015  &   0.352     &   0.015   &  22.9-28.2  &  SW   \\
 NGC 2545  &  260   &   SFD1   &    1.23  &   0.11  &   1.59  &   0.13  &   1.296   &  0.121  &   0.620     &   0.107   &   3.1-6.2   &  W    \\
 NGC 2545  &  260   &   SFD2   &    1.10  &   0.10  &   1.25  &   0.10  &   1.135   &  0.124  &   0.512     &   0.108   &   3.1-7.0   &  E    \\
 NGC 2545  &  260   &   H{\sc ii}     &   13.44  &   0.15  &   5.50  &   0.16  &   0.410   &  0.031  &   0.372     &   0.032   &   7.1-10.6  &  W    \\
 NGC 2523  &   60   &   SFD1   &    ---   &   ---   &   0.34  &   0.04  &    $>$1     &  ---    &   0.306     &   0.147   &   6.2-26.4  &  SW   \\
 NGC 2523  &   60   &   SFD2   &    ---   &   ---   &   ---   &   ---   &    N/A    &  ---    &    N/A      &   ---     &   6.2-21.6  &  NE   \\
 NGC 2523  &   60   &   H{\sc ii}     &   15.35  &   0.21  &   5.86  &   0.10  &   0.382   &  0.022  &   0.351     &   0.023   &  27.7-34.8  &  SW   \\
 NGC 2604  &  320   &   SFD1   &    8.23  &   0.31  &   2.27  &   0.21  &   0.276   &  0.101  &   0.237     &   0.100   &   7.9-13.2  &  NW   \\
 NGC 2604  &  320   &   SFD2   &    6.45  &   0.58  &   1.49  &   0.28  &   0.231   &  0.207  &   0.191     &   0.202   &   8.8-17.6  &  SE   \\
 NGC 2604  &  320   &   H{\sc ii}     &   42.82  &   0.69  &   8.40  &   0.33  &   0.196   &  0.042  &   0.190     &   0.042   &  15.0-19.4  &  NW   \\
 NGC 2604  &  230   &   Bar1   &    4.28  &   0.37  &   1.68  &   0.35  &   0.392   &  0.226  &   0.298     &   0.220   &  13.6-16.3  &  NE   \\
 NGC 2604  &  230   &   Bar2   &   25.23  &   0.38  &   6.96  &   0.25  &   0.276   &  0.039  &   0.262     &   0.039   &  11.9-15.0  &  SW   \\
 NGC 2604  &  230   &   H{\sc ii}     &  121.77  &   0.72  &  22.67  &   0.29  &   0.186   &  0.014  &   0.184     &   0.014   &  15.8-20.7  &  SW   \\
 NGC 2746  &  310   &   SFD1   &    1.44  &   0.10  &   0.75  &   0.17  &   0.523   &  0.234  &   0.271     &   0.232   &   4.4-15.8  &  NW   \\
 NGC 2746  &  310   &   SFD2   &    ---   &   ---   &   0.52  &   0.07  &    $>$1     &  ---  &   0.385     &   0.167   &   5.5-11.7  &  SE   \\
 NGC 2746  &  310   &   H{\sc ii}     &    9.08  &   0.28  &   4.07  &   0.24  &   0.448   &  0.068  &   0.391     &   0.067   &  17.2-24.6  &  NW   \\
 NGC 3049  &  305   &   SFD1   &   12.65  &   0.86  &   7.51  &   0.67  &   0.594   &  0.112  &   0.537     &   0.109   &  12.8-16.3  &  NW   \\
 NGC 3049  &  305   &   SFD2   &    5.05  &   0.70  &   ---  &   ---  &   0.000   &  0.138  &   0.000     &   0.111   &  15.8-29.5  &  SE   \\
 NGC 3049  &  305   &   H{\sc ii}     &   34.33  &   0.73  &   9.91  &   0.58  &   0.289   &  0.063  &   0.278     &   0.063   &  17.6-21.1  &  NW   \\
 NGC 3185  &  300   &   Bar1   &    ---   &   ---   &   1.19  &   0.13  &    $>$1     &  ---  &   0.886     &   0.147   &  7.04-25.1  &  NW   \\
 NGC 3185  &  300   &   Bar2   &    0.62  &   0.10  &   1.80  &   0.19  &   2.907   &  0.199  &   0.918     &   0.136   &  7.04-26.4  &  SE   \\
 NGC 3185  &  300   &   H{\sc ii}     &   18.97  &   0.38  &   7.29  &   0.25  &   0.385   &  0.040  &   0.359     &   0.040   & 33.4-37.8S  &  E    \\
 NGC 3185  &   30   &   SFD1   &    3.02  &   0.35  &   2.75  &   0.40  &   0.910   &  0.187  &   0.630     &   0.170   &  7.04-21.1  &  NE   \\
 NGC 3185  &   30   &   SFD2   &    1.73  &   0.22  &   3.04  &   0.36  &   1.754   &  0.175  &   0.989     &   0.146   &  7.04-19.8  &  SW   \\
 NGC 3185  &   30   &   H{\sc ii}     &   11.90  &   0.37  &   5.33  &   0.38  &   0.448   &  0.078  &   0.402     &   0.077   &  21.1-25.1  &  SW   \\
 NGC 3185  &  315   &   SFD3   &    1.43  &   0.17  &   1.96  &   0.30  &   1.371   &  0.195  &   0.708     &   0.172   &  7.04-31.7  &  NW   \\
 NGC 3185  &  315   &   SFD4   &    0.78  &   0.12  &   2.48  &   0.24  &   3.201   &  0.183  &   1.174     &   0.128   &  7.04-23.3  &  SE   \\
 NGC 3185  &  315   &   H{\sc ii}     &   28.98  &   0.53  &  11.06  &   0.30  &   0.381   &  0.033  &   0.365     &   0.033   & 36.1-40.5   &  SE   \\
 NGC 3185  &  285   &   SFD5   &    1.41  &   0.20  &   1.76  &   0.17  &   1.248   &  0.172  &   0.639     &   0.129   &  7.04-23.8  &  NW   \\
 NGC 3185  &  285   &   SFD6   &    1.14  &   0.15  &   2.16  &   0.16  &   1.888   &  0.148  &   0.870     &   0.109   &  7.04-24.6  &  SE   \\
 NGC 3185  &  285   &   H{\sc ii}     &   13.68  &   0.36  &   5.79  &   0.39  &   0.424   &  0.073  &   0.386     &   0.072   & 29.5-33.4   &  NW   \\
NGC 3346  &    0   &   SFD1   &    2.60  &   0.16  &   1.94  &   0.21  &   0.748   &  0.123  &   0.502     &   0.119   &   6.6-10.6  &  N    \\
 NGC 3346  &    0   &   SFD2   &    2.61  &   0.21  &   1.68  &   0.18  &   0.645   &  0.132  &   0.450     &   0.123   &   5.3-11.4  &  S    \\
 NGC 3346  &    0   &   H{\sc ii}     &   29.37  &   0.23  &   9.09  &   0.19  &   0.309   &  0.022  &   0.296     &   0.023   &  16.3-23.3  &  S    \\
 UGC 5892  &  180   &   SFD1   &    2.51  &   0.23  &   1.87  &   0.18  &   0.747   &  0.131  &   0.487     &   0.117   &   3.5-5.7   &  S    \\
 UGC 5892  &  180   &   SFD2   &    2.40  &   0.10  &   2.00  &   0.22  &   0.836   &  0.118  &   0.536     &   0.119   &   3.5-5.7   &  N    \\
 UGC 5892  &  180   &   H{\sc ii}     &   25.81  &   0.28  &  11.48  &   0.30  &   0.445   &  0.029  &   0.423     &   0.029   &   6.6-9.2   &  N    \\
 NGC 3374  &  285   &   SFD1   &    3.79  &   0.24  &   3.62  &   0.27  &   0.956   &  0.098  &   0.706     &   0.093   &   3.9-6.6   &  ES   \\
 NGC 3374  &  285   &   SFD2   &    4.19  &   0.36  &   2.65  &   0.33  &   0.634   &  0.151  &   0.480     &   0.143   &   4.4-7.5   &  WN   \\
 NGC 3374  &  285   &   H{\sc ii}     &   42.72  &   0.35  &  12.64  &   0.26  &   0.296   &  0.022  &   0.287     &   0.022   &  14.1-20.2  &  ES   \\
\hline
\end{tabular}
\label{tab:EW_ratios}
\end{minipage} 
\end{table*}

\begin{table*}
 \begin{minipage}{140mm}
  \contcaption{}
  \begin{tabular}{lrlcccccccccc}
  \hline
  Name   & Slit   &   Region &   EW$_{\rm{H\alpha}}$   &   err   &  EW$_{\rm{[N{\sc ii}]}}$  &   err   & [N{\sc ii}]/H$\alpha$      &  err      & [N{\sc ii}]/H$\alpha$      &     err   &  Rad range  & Dir  \\
         &  ~PA   &          &          &         &         &         & ~(obs)      &           & ~(cor)      &           &      $^{\prime\prime}$        &       \\
  \hline
 NGC 3485  &  130   &   SFD1   &    ---   &   ---   &   1.08  &   0.16  &    $>$1     &  ---  &   1.078     &   0.181   &   5.3-14.5  &  NW   \\
 NGC 3485  &  130   &   SFD2   &    ---   &   ---   &   0.94  &   0.10  &    $>$1     &  ---  &   0.609     &   0.149   &   5.3-12.8  &  SE   \\
 NGC 3485  &  130   &   H{\sc ii}     &   48.79  &   0.33  &  16.26  &   0.24  &   0.333   &  0.016  &   0.324     &   0.016   &  24.6-30.0  &  NW   \\
 NGC 3507  &   30   &   SFD1   &    1.10  &   0.11  &   1.57  &   0.12  &   1.434   &  0.124  &   0.635     &   0.105   &   6.2-23.8  &  NE   \\
 NGC 3507  &   30   &   SFD2   &    0.94  &   0.09  &   1.31  &   0.09  &   1.389   &  0.112  &   0.563     &   0.096   &   6.2-26.4  &  SW   \\
 NGC 3507  &   30   &   H{\sc ii}     &   20.35  &   0.19  &   7.97  &   0.19  &   0.392   &  0.025  &   0.367     &   0.026   &  36.1-40.5  &  NE   \\
 NGC 3729  &  290   &   SFD1   &    4.42  &   0.13  &   4.70  &   0.19  &   1.062   &  0.050  &   0.820     &   0.051   &   4.4-14.1  &  NW   \\
 NGC 3729  &  290   &   SFD2   &    3.91  &   0.13  &   3.69  &   0.23  &   0.942   &  0.071  &   0.695     &   0.072   &   6.2-14.5  &  SE   \\
 NGC 3729  &  290   &   Clump   &    5.39  &   0.16  &   5.63  &   0.19  &   1.045   &  0.045  &   0.837     &   0.046   &   4.4-7.0   &  NW   \\
 NGC 3729  &  290   &   H{\sc ii}     &   29.38  &   0.12  &   9.77  &   0.14  &   0.332   &  0.015  &   0.318     &   0.016   &  17.2-22.4  &  SE   \\
 NGC 3963  &  220   &   SFD   &    2.00  &   0.19  &   1.62  &   0.16  &   0.810   &  0.138  &   0.485     &   0.123   &   5.7-17.2  &  SW   \\
 NGC 3963  &  220   &   H{\sc ii}     &   19.91  &   0.25  &   7.65  &   0.23  &   0.384   &  0.033  &   0.360     &   0.033   &  19.8-26.0  &  SW   \\
 NGC 4123  &   15   &   SFD1   &    4.72  &   0.39  &   3.80  &   0.27  &   0.805   &  0.110  &   0.612     &   0.099   &   8.8-26.4  &  N    \\
 NGC 4123  &   15   &   SFD2   &    3.13  &   0.37  &   3.70  &   0.44  &   1.182   &  0.166  &   0.828     &   0.147   &   8.8-23.3  &  S    \\
 NGC 4123  &   15   &   H{\sc ii}     &   32.75  &   0.99  &  10.08  &   0.64  &   0.308   &  0.071  &   0.296     &   0.070   &  42.7-46.2  &  N    \\
 NGC 4416  &   95   &   SFD1   &   10.69  &   0.25  &   3.36  &   0.16  &   0.315   &  0.053  &   0.280     &   0.053   &   4.4-8.8   &  E    \\
 NGC 4416  &   95   &   SFD2   &   11.04  &   0.23  &   4.16  &   0.22  &   0.377   &  0.056  &   0.336     &   0.057   &   4.4-8.8   &  W    \\
 NGC 4416  &   95   &   H{\sc ii}     &   23.40  &   0.46  &   6.88  &   0.47  &   0.294   &  0.071  &   0.278     &   0.071   &  22.9-26.8  &  W    \\
 NGC 4619  &  100   &   SFD1   &    2.01  &   0.06  &   2.03  &   0.12  &   1.009   &  0.067  &   0.606     &   0.074   &   2.6-6.2   &  E    \\
 NGC 4619  &  100   &   SFD2   &    2.05  &   0.12  &   1.97  &   0.16  &   0.958   &  0.099  &   0.580     &   0.096   &   2.6-5.3   &  W    \\
 NGC 4619  &  100   &   H{\sc ii}     &   14.35  &   0.12  &   6.46  &   0.17  &   0.450   &  0.028  &   0.412     &   0.029   &  7.0-11.0   &  W    \\
 NGC 4779  &  100   &   SFD1   &    3.83  &   0.30  &   2.99  &   0.16  &   0.781   &  0.096  &   0.569     &   0.084   &   5.3-13.6  &  E    \\
 NGC 4779  &  100   &   SFD2   &    2.45  &   0.20  &   2.08  &   0.14  &   0.849   &  0.104  &   0.558     &   0.091   &   5.3-18.0  &  W    \\
 NGC 4779  &  100   &   H{\sc ii}     &  193.95  &   0.83  &  58.55  &   0.30  &   0.302   &  0.007  &   0.300     &   0.007   &  20.2-24.6  &  W    \\
 NGC 4904  &   45   &   SFD1   &    3.90  &   0.19  &   1.96  &   0.10  &   0.502   &  0.070  &   0.373     &   0.067   &  11.4-22.4  &  NE   \\
 NGC 4904  &   45   &   H{\sc ii}     &   53.41  &   0.58  &  13.11  &   0.26  &   0.245   &  0.023  &   0.239     &   0.023   &  30.4-24.8  &  SW   \\
 NGC 4904  &  315   &   Bar1   &    9.36  &   0.26  &   3.59  &   0.16  &   0.384   &  0.051  &   0.336     &   0.051   &  19.4-21.6  &  NW   \\
 NGC 4904  &  315   &   H{\sc ii}     &   50.69  &   0.75  &  13.18  &   0.16  &   0.260   &  0.019  &   0.253     &   0.019   &  23.8-27.7  &  NW   \\
 NGC 4999  &  330   &   SFD1   &    1.45  &   0.08  &   1.47  &   0.16  &   1.012   &  0.122  &   0.520     &   0.121   &   6.2-16.3  &  NW   \\
 NGC 4999  &  330   &   SFD2   &    1.65  &   0.31  &   1.01  &   0.03  &   0.608   &  0.191  &   0.310     &   0.113   &   6.2-20.7  &  SE   \\
 NGC 4999  &  330   &   H{\sc ii}     &   10.88  &   0.18  &   3.12  &   0.14  &   0.287   &  0.048  &   0.255     &   0.048   &  22.4-26.4  &  SE   \\
 NGC 5164  &  115   &   SFD1   &    1.52  &   0.18  &   1.31  &   0.27  &   0.863   &  0.239  &   0.458     &   0.223   &   3.5-7.1   &  SE   \\
 NGC 5164  &  115   &   SFD2   &    ---   &   ---   &   ---   &   ---   &    N/A    &  ---  &    N/A      &    ---    &   3.5-6.6   &  NW   \\
 NGC 5164  &  115   &   H{\sc ii}     &   11.70  &   0.49  &   4.01  &   0.30  &   0.343   &  0.085  &   0.308     &   0.084   &  9.7-13/2   &  SE   \\
 NGC 5350  &   20   &   SFD1   &    0.80  &   0.13  &   1.12  &   0.14  &   1.402   &  0.207  &   0.534     &   0.156   &   6.2-13.6  &  N    \\
 NGC 5350  &   20   &   SFD2   &    ---   &   ---   &   1.57  &   0.15  &    $>$1     &  ---  &   1.210     &   0.140   &   6.6-16.7  &  S    \\
 NGC 5350  &   20   &   H{\sc ii}     &   35.45  &   0.16  &  12.61  &   0.22  &   0.356   &  0.018  &   0.343     &   0.018   &  32.6-39.6  &  S    \\
 NGC 5375  &  260   &   SFD1   &    0.67  &   0.11  &   0.67  &   0.08  &   1.010   &  0.200  &   0.330     &   0.147   &   4.4-19.4  &  W    \\
 NGC 5375  &  260   &   SFD2   &    ---   &   ---   &   0.46  &   0.04  &    $>$1     &  ---  &   0.310     &   0.129   &   4.4-20.2  &  E    \\
 NGC 5375  &  260   &   H{\sc ii}     &    6.46  &   0.23  &   3.11  &   0.26  &   0.481   &  0.090  &   0.399     &   0.090   &  22.9-30.3  &  W    \\
 NGC 5618  &   95   &   SFD1   &    1.65  &   0.34  &   ---   &   ---   &   0.000   &  0.208  &   0.000     &   0.123   &   3.5-7.0   &  E    \\
 NGC 5618  &   95   &   SFD2   &    1.30  &   0.31  &   0.96  &   0.21  &   0.741   &  0.329  &   0.364     &   0.257   &   3.5-6.6   &  W    \\
 NGC 5618  &   95   &   H{\sc ii}     &    8.34  &   0.26  &   4.94  &   0.45  &   0.592   &  0.095  &   0.510     &   0.095   &  12.8-16.3  &  W    \\
 IC 1010   &   80   &   SFD1   &    ---   &   ---   &   1.98  &   0.19  &    $>$1     &  ---  &   1.472     &   0.140   &   4.8-11.9  &  E    \\
 IC 1010   &   80   &   SFD2   &    ---   &   ---   &   ---   &   ---   &    N/A    &  ---  &    N/A      &   ---     &   4.8-8.4   &  W    \\
 IC 1010   &   80   &   H{\sc ii}     &   14.65  &   0.41  &   7.06  &   0.45  &   0.482   &  0.070  &   0.441     &   0.069   &  13.2-17.2  &  W    \\
 NGC 5735  &  270   &   SFD1   &    1.30  &   0.16  &   1.17  &   0.15  &   0.904   &  0.178  &   0.445     &   0.149   &   6.6-11.4  &  N    \\
 NGC 5735  &  270   &   SFD2   &    0.88  &   0.13  &   2.04  &   0.21  &   2.307   &  0.182  &   0.917     &   0.133   &   5.3-11.0  &  S    \\
 NGC 5735  &  270   &   H{\sc ii}     &   14.70  &   0.33  &   5.50  &   0.20  &   0.374   &  0.043  &   0.343     &   0.043   &  22.4-27.3  &  S    \\
 NGC 5735  &    0   &   Bar1   &    1.56  &   0.20  &   1.90  &   0.21  &   1.222   &  0.171  &   0.657     &   0.140   &   3.1-6.2   &  W    \\
 NGC 5735  &    0   &   Bar2   &    2.36  &   0.09  &   1.50  &   0.14  &   0.634   &  0.101  &   0.404     &   0.103   &   2.6-6.1   &  E    \\
 NGC 5735  &    0   &   H{\sc ii}     &   22.51  &   0.19  &   7.73  &   0.18  &   0.343   &  0.025  &   0.324     &   0.026   &  10.1-14.1  &  E    \\
 IC 1067   &   45   &   SFD1   &    ---   &   ---   &   0.64  &   0.11  &    $>$1     &  ---  &   0.481     &   0.195   &   3.9-12.8  &  NE   \\
 IC 1067   &   45   &   SFD2   &    ---   &   ---   &   0.55  &   0.12  &    $>$1     &  ---  &   0.410     &   0.243   &   4.4-12.3  &  SW   \\
 IC 1067   &   45   &   H{\sc ii}     &   26.66  &   0.51  &  11.34  &   0.51  &   0.425   &  0.049  &   0.405     &   0.049   &  14.5-19.8  &  SW   \\
 NGC 5970  &  350   &   SFD1   &    0.81  &   0.07  &   1.23  &   0.08  &   1.521   &  0.108  &   0.539     &   0.095   &   4.4-8.8   &  N    \\
 NGC 5970  &  350   &   SFD2   &    1.20  &   0.05  &   1.42  &   0.07  &   1.181   &  0.063  &   0.497     &   0.078   &   4.4-9.7   &  S    \\
 NGC 5970  &  350   &   H{\sc ii}     &   74.01  &   1.04  &  28.35  &   0.44  &   0.383   &  0.021  &   0.376     &   0.021   &  15.4-20.2  &  N    \\
 NGC 5970  &   80   &   Bar1   &    0.84  &   0.13  &   1.12  &   0.08  &   1.338   &  0.172  &   0.515     &   0.111   &   3.1-7.5   &  E    \\
 NGC 5970  &   80   &   Bar2   &    1.78  &   0.08  &   1.29  &   0.05  &   0.724   &  0.059  &   0.413     &   0.063   &   3.9-12.8  &  W    \\
\hline
\end{tabular}
\label{tab:EW_ratios}
\end{minipage} 
\end{table*}

\begin{table*}
 \begin{minipage}{140mm}
  \contcaption{}
  \begin{tabular}{lrlcccccccccc}
  \hline
  Name   & Slit   &   Region &   EW$_{\rm{H\alpha}}$   &   err   &  EW$_{\rm{[N{\sc ii}]}}$  &   err   & [N{\sc ii}]/H$\alpha$      &  err      & [N{\sc ii}]/H$\alpha$      &     err   &  Rad range  & Dir  \\
         &  ~PA   &          &          &         &         &         & ~(obs)      &           & ~(cor)      &           &      $^{\prime\prime}$        &       \\
  \hline
 NGC 5970  &   80   &   H{\sc ii}     &   65.87  &   0.89  &  20.37  &   0.41  &   0.309   &  0.024  &   0.303     &   0.024   &  26.4-30.4  &  E    \\
 NGC 5970  &   65   &   SFD3   &    0.79  &   0.10  &   0.82  &   0.05  &   1.036   &  0.140  &   0.384     &   0.097   &   4.0-9.7   &  EN   \\
 NGC 5970  &   65   &   SFD4   &    1.03  &   0.14  &   0.92  &   0.06  &   0.897   &  0.149  &   0.389     &   0.102   &   4.0-8.4   &  WS   \\
 NGC 5970  &   65   &   H{\sc ii}     &   13.21  &   0.15  &   4.36  &   0.16  &   0.330   &  0.039  &   0.300     &   0.040   &  14.5-18.9  &  EN   \\
 NGC 5970  &   95   &   SFD5   &    1.63  &   0.14  &   1.02  &   0.08  &   0.630   &  0.114  &   0.345     &   0.099   &   3.5-8.4   &  E    \\
 NGC 5970  &   95   &   SFD6   &    2.09  &   0.13  &   1.29  &   0.06  &   0.616   &  0.079  &   0.376     &   0.072   &   2.6-7.0   &  W    \\
 NGC 5970  &   95   &   H{\sc ii}     &   20.52  &   0.14  &   5.15  &   0.11  &   0.251   &  0.023  &   0.236     &   0.024   &  15.4-9.8   &  E    \\
 NGC 6004  &  100   &   SFD   &    2.35  &   0.31  &   2.17  &   0.17  &   0.925   &  0.155  &   0.589     &   0.122   &   3.5-7.9   &  E    \\
 NGC 6004  &  100   &   H{\sc ii}     &   23.52  &   0.30  &   8.64  &   0.30  &   0.367   &  0.037  &   0.348     &   0.037   &  13.6-17.1  &  E    \\
  \hline
 NGC  864  &   55   &    SFD1  &    ---   &   ---   &   0.88  &   0.16  &   $>$1      &  ---  &   0.656     &  0.207    &  10.12-32.12  &  NE    \\
 NGC  864  &   55   &    SFD2  &    1.31  &   0.16  &   ---   &   ---   &   0.000   &  ---  &   0.000     &  1.003    &  19.36-33.88  &  SW    \\
 NGC  864  &   55   &    H{\sc ii}    &   93.18  &   0.93  &  20.81  &   0.84  &   0.223   &  0.040  &   0.220     &  0.042    &  67.76-73.92  &  NE    \\
 NGC 2268  &  183   &    SFD1  &    1.94  &   0.12  &   2.46  &   0.16  &   1.267   &  0.069  &   0.750     &  0.085    &   3.08-4.84   &  N     \\
 NGC 2268  &  183   &    SFD2  &    2.81  &   0.09  &   2.85  &   0.13  &   1.017   &  0.047  &   0.687     &  0.060    &   3.96-7.04   &  S     \\
 NGC 2268  &  183   &    H{\sc ii}    &   72.14  &   1.03  &  25.32  &   0.51  &   0.351   &  0.020  &   0.345     &  0.025    &   8.36-11.44  &  S     \\
 UGC 3685  &    0   &    SFD1  &    0.83  &   0.15  &   1.72  &   0.28  &   2.076   &  0.195  &   0.942     &  0.190    &   7.48-17.60  &  N     \\
 UGC 3685  &    0   &    SFD2  &    1.60  &   0.13  &   1.44  &   0.21  &   0.895   &  0.152  &   0.510     &  0.159    &   7.48-17.60  &  S     \\
 UGC 3685  &    0   &    H{\sc ii}    &   48.09  &   0.69  &  13.63  &   0.45  &   0.283   &  0.033  &   0.276     &  0.036    &  30.80-36.08  &  S     \\
 NGC 2543  &  180   &    SFD1  &    1.19  &   0.21  &   1.17  &   0.31  &   0.983   &  0.296  &   0.402     &  0.281    &   7.48-24.20  &  N     \\
 NGC 2543  &  180   &    SFD2  &    1.67  &   0.20  &   1.88  &   0.37  &   1.125   &  0.211  &   0.557     &  0.212    &   6.60-19.80  &  S     \\
 NGC 2543  &  180   &    H{\sc ii}    &   86.41  &   1.73  &  26.75  &   0.77  &   0.310   &  0.029  &   0.305     &  0.035    &  23.32-26.40  &  S     \\
 NGC 2595  &  260   &    SFD1  &    ---   &   ---   &   ---   &   ---   &    NA     &  ---  &    N/A      &  ---    &   9.68-19.36  &  E     \\
 NGC 2595  &  260   &    SFD2  &    2.59  &   0.29  &   1.79  &   0.30  &   0.693   &  0.180  &   0.416     &  0.185    &  13.20-26.40  &  W     \\
 NGC 2595  &  260   &    H{\sc ii}    &   26.31  &   0.34  &  10.72  &   0.25  &   0.408   &  0.023  &   0.388     &  0.027    &  23.76-29.92  &  E     \\
 NGC 2712  &  350   &    SFD1  &    1.00  &   0.11  &   1.44  &   0.16  &   1.441   &  0.123  &   0.618     &  0.134    &   6.16-21.56  &  S     \\
 NGC 2712  &  350   &    SFD2  &    1.49  &   0.10  &   1.59  &   0.15  &   1.069   &  0.099  &   0.577     &  0.111    &   7.48-20.68  &  N     \\
 NGC 2712  &  350   &    H{\sc ii}    &   19.66  &   0.33  &   7.58  &   0.23  &   0.385   &  0.031  &   0.361     &  0.035    &  33.00-44.44  &  N     \\
 NGC 3185  &  320   &    SFD1  &    0.39  &   0.13  &   1.14  &   0.11  &   2.908   &  0.208  &   0.728     &  0.148    &   9.68-21.56  &  SE    \\
 NGC 3185  &  320   &    SFD2  &    0.45  &   0.11  &   0.84  &   0.12  &   1.862   &  0.203  &   0.552     &  0.175    &  11.44-29.04  &  NW    \\
 NGC 3185  &  320   &    H{\sc ii}    &   10.83  &   0.21  &   4.34  &   0.18  &   0.401   &  0.042  &   0.357     &  0.046    &  33.44-42.68  &  NW    \\
 NGC 3351  &  200   &    SFD1  &    ---   &   ---   &   1.95  &   0.42  &   $>$1      &  ---  &   1.454     &  0.237    &  27.28-51.48  &  NNE   \\
 NGC 3351  &  200   &    SFD2  &    ---   &   ---   &   1.33  &   0.13  &   $>$1      &  ---  &   0.992     &  0.140    &  25.52-40.92  &  SSW   \\
 NGC 3351  &  200   &    H{\sc ii}    &   16.37  &   0.40  &   4.99  &   0.35  &   0.305   &  0.071  &   0.282     &  0.074    &  73.48-58.08  &  NNE   \\
 NGC 3367  &  170   &    SFD1  &    2.39  &   0.17  &   3.03  &   0.07  &   1.271   &  0.028  &   0.848     &  0.062    &   5.28-9.68   &  N     \\
 NGC 3367  &  170   &    SFD2  &    4.76  &   0.36  &   5.27  &   0.29  &   1.107   &  0.061  &   0.898     &  0.085    &   5.72-10.12  &  S     \\
 NGC 3367  &  170   &    H{\sc ii}    &   46.25  &   0.58  &  13.55  &   0.25  &   0.293   &  0.019  &   0.285     &  0.022    &  11.88-16.28  &  N     \\
 NGC 3811  &  267   &    SFD1  &    1.13  &   0.11  &   1.07  &   0.13  &   0.950   &  0.131  &   0.426     &  0.140    &   5.28-10.56  &  W     \\
 NGC 3811  &  267   &    SFD2  &    0.35  &   0.10  &   1.24  &   0.13  &   3.581   &  0.186  &   0.750     &  0.144    &   5.28-10.56  &  E     \\
 NGC 3811  &  267   &    H{\sc ii}    &  103.17  &   2.43  &  33.05  &   0.92  &   0.320   &  0.028  &   0.316     &  0.036    &  21.56-24.64  &  W     \\
 NGC 4051  &  117   &    SFD   &    0.19  &   0.04  &   0.56  &   0.04  &   2.897   &  0.116  &   0.366     &  0.116    &  12.76-32.12  &  NW    \\
 NGC 4051  &  117   &    H{\sc ii}    &   34.88  &   0.92  &  11.49  &   0.35  &   0.329   &  0.031  &   0.317     &  0.040    &  71.28-81.84  &  NW    \\
 NGC 4210  &  170   &    SFD1  &    0.65  &   0.11  &   1.30  &   0.11  &   2.017   &  0.113  &   0.653     &  0.121    &   3.08-5.72   &  N     \\
 NGC 4210  &  170   &    SFD2  &    ---   &   ---   &   1.07  &   0.14  &   $>$1      &  ---  &   0.798     &  0.165    &   3.52-6.16   &  S     \\
 NGC 4210  &  170   &    H{\sc ii}    &   15.99  &   0.08  &   6.15  &   0.20  &   0.385   &  0.033  &   0.355     &  0.034    &  11.00-15.40  &  N     \\
 NGC 5698  &   70   &    SFD1  &    1.05  &   0.25  &   1.93  &   0.24  &   1.831   &  0.181  &   0.852     &  0.175    &   6.16-22.00  &  WSW   \\
 NGC 5698  &   70   &    SFD2  &    ---   &   ---   &   2.64  &   0.20  &   $>$1      &  ---  &   1.992     &  0.125    &   7.04-19.36  &  ENE   \\
 NGC 5698  &   70   &    H{\sc ii}    &   74.92  &   0.63  &  24.39  &   0.56  &   0.326   &  0.023  &   0.320     &  0.024    &  33.88-37.84  &  ENE   \\
 NGC 5806  &   30   &    SFD1  &    0.84  &   0.07  &   1.74  &   0.13  &   2.072   &  0.082  &   0.794     &  0.102    &   6.60-13.20  &  SW    \\
 NGC 5806  &   30   &    SFD2  &    0.60  &   0.09  &   1.65  &   0.12  &   2.744   &  0.095  &   0.892     &  0.111    &   6.60-11.44  &  NE    \\
 NGC 5806  &   30   &    H{\sc ii}    &    6.65  &   0.16  &   2.92  &   0.14  &   0.440   &  0.049  &   0.365     &  0.055    &  15.40-24.20  &  NE    \\
UGC 10888  &  120   &    SFD1  &    0.89  &   0.16  &   2.08  &   0.38  &   2.332   &  0.215  &   0.933     &  0.205    &   3.96-9.24   &  NW    \\
UGC 10888  &  120   &    SFD2  &    ---   &   ---   &   0.92  &   0.15  &   $>$1      &  ---  &   0.686     &  0.191    &   3.96-8.80   &  SE    \\
UGC 10888  &  120   &    H{\sc ii}    &   22.08  &   0.32  &   9.42  &   0.29  &   0.427   &  0.031  &   0.402     &  0.034    &  10.56-18.48  &  NW    \\
\hline
\end{tabular}
\label{tab:EW_ratios}
\end{minipage} 
\end{table*}

\subsection{Data reduction}

Data reduction of our long-slit spectroscopy was performed using Starlink software.  All stages, i.e. bias subtraction, flat fielding, correction of spectra for minor rotation and spatial distortion effects, sky background subtraction, wavelength calibration using a copper neon$+$ copper argon arc lamp and flux calibration from spectrophotometric standard observations were completely standard are will not be described further here.  More details are given in \cite{jame15}, which also describes the use of H$\alpha$ narrow-band imaging to define the apparently emission-line free region from which the SFD spectra were extracted. In most cases, two SFD regions were extracted for each galaxy, one on either side of the nucleus (taking care to exclude all emission from the nuclear region). 

The new analysis presented here resulted in 72 SFD spectra, from 34 galaxies, from which we were able to determine [N{\sc ii}]/H$\alpha$ emission line ratios for 66 of these regions (4 regions had no detected line emission, and 2 more showed H$\alpha$ but no [N{\sc ii}]). In combination with the 29 regions in 15 galaxies discussed in \cite{jame15} and reanalysed here, this gives us a sample of 101 observed  regions from a sample of 48 galaxies (noting that NGC\,3185 was included in our earlier study and re-observed here), 93 of which have line detections enabling the calculation of line ratios. We also observed 6 galaxies with the slit aligned with the major axis of the bar, resulting in 11 off-nucleus bar regions with measured spectroscopic parameters to enable a preliminary comparison of bar and SFD properties.

As previously, at least one region of strong line emission from an H{\sc ii} region well outside the `desert' was defined for all of the  galaxies, to give reference line ratios corresponding to star formation.  Line ratios were extracted for 59 such H{\sc ii} regions in the 48 galaxies of the extended sample.

This analysis resulted in line fluxes, line widths and central wavelengths for the H$\alpha$ 6563\,\AA\ line, and the same parameters for the stronger of the two [N{\sc ii}] lines at 6584\,\AA\ plus equivalent widths for both lines.

\section{Results}

\subsection{Observed emission-line ratios}

\begin{figure}
\includegraphics[width=90mm,angle=0]{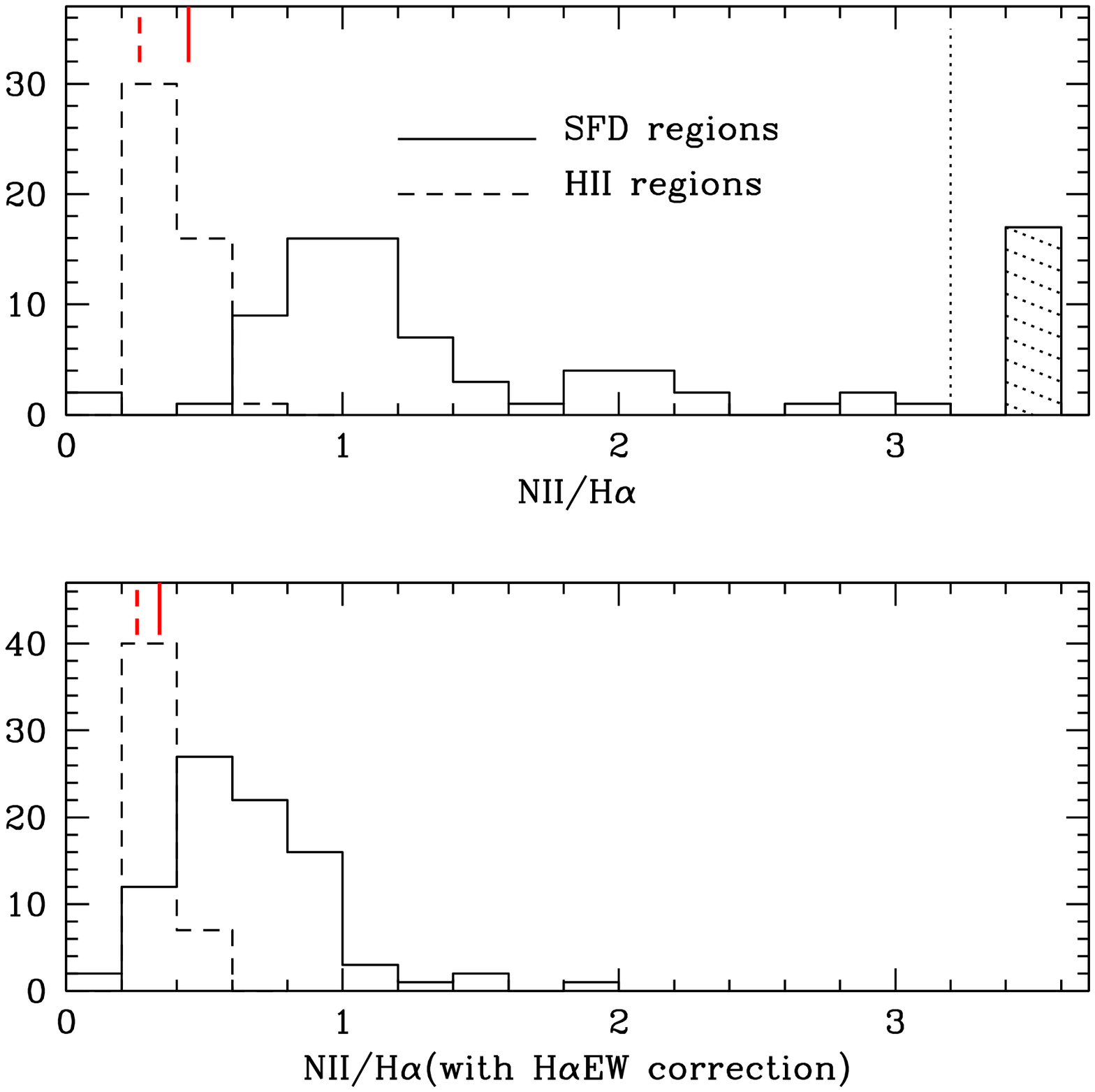}
\caption{ Upper frame: Histogram of observed [N{\sc ii}]/H$\alpha$ line ratios for SFD (solid line) and H{\sc ii} (dashed line) regions with data presented here. The H$\alpha$ fluxes have no correction for underlying H$\alpha$ absorption features. The points to the right of the dotted line correspond to regions detected in [N{\sc ii}] only. Only SFD regions observed with the slit perpendicular to the bar axis (`SFD1' and `SFD2' in Table 2) are included.  Lower frame: As for the upper frame, but after correction for underlying H$\alpha$ absorption. The solid and dashed lines to the upper left show the mean line ratios for equivalent regions in 9 SFD regions within 5 late-type Scd spiral galaxies. }
\label{fig:ratios_hist}
\end{figure}

The upper frame in Fig.~\ref{fig:ratios_hist} shows the distribution of values of the observed
[N{\sc ii}]6584\,\AA/H$\alpha$6563\,\AA\ line ratio (henceforth [N{\sc ii}]/H$\alpha$), for all of the SFD regions for which the two lines were detected.   Ratios derived from SFD regions are shown as the solid
histogram, while the corresponding values for the comparison H{\sc ii} regions are shown
using dashed lines. The main histogram does not include line ratio data for the 5 late-type SBcd galaxies, however.  While measurements were made for 9 regions in these galaxies, that are listed as `SFD' in Table~\ref{tab:EW_ratios}, all 9 show evidence for ongoing star formation, and hence in this plot we simply show their mean line ratio as a  solid line in the upper-left hand corner of the plot, to demonstrate their consistency with the ratios found for H{\sc ii} regions (dashed histogram).  
As was found in \cite{jame15}, the latter form a very tight distribution around the values typical of excitation by star formation as found by numerous 
authors following the work of \citet{bald81}.

In the upper frame in Fig.~\ref{fig:ratios_hist}, 17 regions are plotted in a single shaded histogram box to far right of the plot. These are regions which only had clear detections of the [N{\sc ii}] line, and no H$\alpha$, in the initial reduction of these spectra.  However, these spectra still contain stellar absorption features, and in the next subsection we show that correction for absorption lines does reveal the presence of line emission in all of these cases.

\subsection{Correction for underlying H$\alpha$ absorption}

Measured H$\alpha$ emission-line fluxes will be reduced due to the presence of H$\alpha$ absorption features. In \cite{jame15}, we corrected for this by assuming that an absorption feature of the same equivalent width (EW) was present in all galaxy spectra.  The EW value used was 1.103\,\AA, with an adopted uncertainty of 0.210\,\AA, within a bandpass of 8\,\AA.  This corresponded to a simple stellar population (SSP) with a solar metallicity and an age of 3.5\,Gyr, taken from the BaSTI population synthesis database \citep{perc09}.  The BaSTI spectrum corresponding to this SSP was a reasonable `by-eye' fit to the SFD spectra for the 15 galaxies discussed in \cite{jame15}.  

However, a more sophisticated and self-consistent correction was needed for subsequent papers \citep{jame16,jame18}, where the exact strength of the H$\beta$ absorption feature was needed, and infilling by H$\beta$ emission could strongly bias the stellar ages resulting from this measurement.  Hence we adopted an iterative method enabling simultaneous fitting of emission and absorption features, which is described fully in \cite{jame16}. Briefly, this procedure starts by determining the strength of the H$\beta$ absorption feature by direct integration on the observed spectrum.  This value is then used to identify the age of the best-fitting BaSTI SSP.  The H$\alpha$ absorption EW is then taken from this model spectrum, and used to correct the H$\alpha$ emission line flux in the observed spectrum.  This is turn is used to predict the H$\beta$ emission-line strength, using an H$\alpha$/H$\beta$ flux ratio of 2.85 \citep{broc71,humm87}.  The initially-measured value of the H$\beta$ absorption is then corrected for the assumed presence of this emission line, and the corrected value is used to start the next stage of iterations.  

In all cases, 3 or 4 cycles of iteration were sufficient to find a stable, self-consistent set of values for the H$\alpha$ and H$\beta$ emission and absorption line strengths.  In \cite{jame18}, this process was applied to spectra of SFD regions in 21 galaxies, all of which are in the present study.  A further 6 of the present sample have spectroscopy covering the H$\alpha$, H$\beta$ and Mgb lines, enabling the implementation of this iterative method.  For the remaining 21 galaxies, only red spectroscopy covering the H$\alpha$ and [N{\sc ii}] regions was available.  For these, the H$\alpha$ fluxes have been corrected assuming a fixed absorption EW.  However, the best-fit models for the iterative fits to spectra from the other 27 galaxies corresponded to SSPs with significantly younger ages, 1.9\,Gyr rather than the 3.5\,Gyr assumed in \cite{jame15}.  This results in a correction with 1.34\,\AA\ EW, rather than the 1.103\,\AA\ EW used previously.  The new value is applied to all of the spectra for which the iterative method cannot be used, and hence updated values of [N{\sc ii}]/H$\alpha$ are presented in Table 2 (below the horizontal line) for the spectroscopy already presented in \cite{jame15}. The updated corrections will be used for all line ratios and H$\alpha$ EW values discussed in the remainder of this paper, for SFD, H{\sc ii} and bar regions.

\subsection{Emission-line ratios and equivalent widths after absorption correction}

The lower frame of Fig.~\ref{fig:ratios_hist} shows the effect on [N{\sc ii}]/H$\alpha$ line ratios of boosting the H$\alpha$ emission-line
fluxes using either the iterative correction method or the 1.34~\AA\ equivalent width correction as explained above.  The effect on ratios from H{\sc ii} regions is small, as expected given the high EW line emission.  For the SFD regions, the absorption line correction significantly compresses the range of measured ratios, and removes the group of 17 regions for which only [N{\sc ii}] emission was detected.  After application of the absorption correction, all are concluded to have H$\alpha$ emission, at levels that make their line ratios consistent with the remainder of the population.  
More quantitatively, for the 87 SFD regions, after the application of absorption corrections and with the late-type Scd galaxies removed, the mean observed [N{\sc ii}]/H$\alpha$ ratio is 0.648$\pm$0.033, median
0.573.  
For the 53 H{\sc ii} regions in the same galaxies, the mean 
absorption-corrected [N{\sc ii}]/H$\alpha$
ratio is 0.347$\pm$0.008, median 0.355.  A Kolmogorov-Smirnov test applied to the distributions of SFD and H{\sc ii} region line ratios gives a maximum difference in the cumulative normalised distributions $D=$0.7554, with a probability $P=$0.000 of their being drawn from the same parent distribution. Thus, while the strong absorption correction has closed up the distributions, the SFD region ratios are much too high to be consistent with SF.

The mean corrected line ratio for SFD regions in late-type Scd galaxies is now 0.338$\pm$0.058, more consistent with the values for H{\sc ii} regions than for SFD regions of galaxies of other, earlier types. In these 5 galaxies only, the line emission in the SFD regions {\em does} seem to be dominated by SF. This confirms the suggestion by \cite{nair10} and \cite{hako14} that bars in late-type disc galaxies have quite different effects on star formation properties from those with earlier-type hosts, and justifies our separating out these 5 galaxies from the remainder of the sample in our analysis of the SFD phenomenon in bar-swept regions.

\begin{figure}
\includegraphics[width=90mm,angle=0]{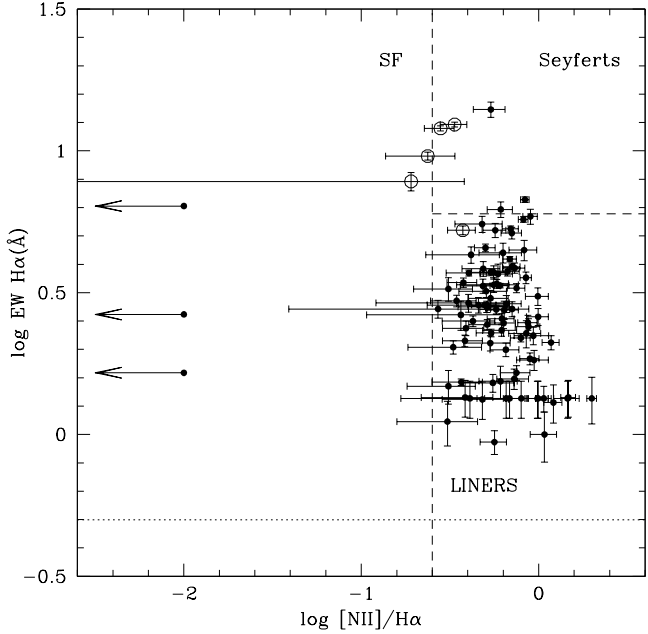}
\caption{A `WHAN' diagram \citep{cidf11} showing H$\alpha$ equivalent width as a function of [N{\sc ii}]/H$\alpha$ line ratio, where the dashed lines mark the boundaries separating emission powered by star formation, Seyfert/AGN activity and LINER-type emission.  The dotted line corresponds to an H$\alpha$ equivalent width of 0.5 \AA, the upper limit proposed in the literature for emission from old populations in early-type galaxies.
}
\label{fig:WHAN}
\end{figure}

We are now in a position to analyse H$\alpha$ and [N{\sc ii}] EW values and line ratios for the SFD regions, with the H$\alpha$ values having been corrected for absorption. The mean EW values are 2.791$\pm$0.134\,\AA\ for H$\alpha$, and 1.762$\pm$0.106\,\AA\ for [N{\sc ii}].  These means are somewhat higher than the predictions for fully passive galaxies from the models of \cite{byle17} and \cite{byle19}, and than the observed EW values for the `retired galaxies' of \cite{cidf11}. Figure~\ref{fig:WHAN} shows the distribution of the SFD regions on a `WHAN' diagram \citep{cidf11} of H$\alpha$ equivalent-width as a function of [N{\sc ii}]/H$\alpha$ line ratio. Almost all of the SFD regions lie within the locus defined by \cite{cidf11} to correspond to LINER-type emission. A small number of regions have sufficiently strong emission lines that they cross into the Seyfert/AGN or star formation region; these include most of the galaxies that we identified above as being late-type galaxies with ongoing SF (open circles in Fig.~\ref{fig:WHAN}). The arrows correspond to SFD spectra where H$\alpha$ lines are measurable but no [N{\sc ii}] lines are seen. These results will be discussed further in Sect. 4.

\subsection{Is this diffuse emission excited directly by bars?}

One possible mechanism for exciting line emission with high [N{\sc ii}]/H$\alpha$ ratios is through shocks \citep{bald81}, given that the shock models of e.g. \cite{alle08} nicely reproduce the [N{\sc ii}]/H$\alpha$ line ratios found here, and in SFD regions shocks could plausibly be driven by the bar itself.  In order to test this possibility, we performed an analysis of two galaxies using multiple slit angles, to test whether emission peaks `downstream' of bars.  The two galaxies chosen were NGC\,3185 and NGC\,5970, types SBa and SBc respectively, both of which show `normal' LINER-type emission in their SFD regions as measured perpendicular to the bar axis. Both were additionally observed with the slit position angle along the bar,  at $+15^{\circ}$ and $-15^{\circ}$ to the bar major axis, and also at $+20^{\circ}$ for NGC\,3185 (positive angles are defined as increasing anticlockwise, which is the rotation direction deduced for both galaxies, assuming trailing spiral arms). The resulting values are listed in Table~\ref{tab:EW_ratios} as `SFD3' - `SFD6'.

\begin{figure}
\includegraphics[width=90mm,angle=0]{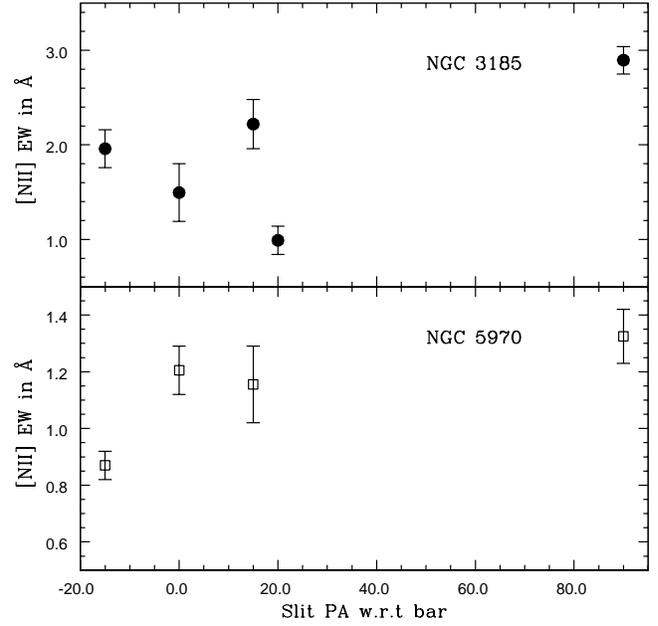}
\caption{
[N{\sc ii}] equivalent widths for multiple position angles relative to the bar axis.  Solid circles show values for NGC\,3185, and open squares the ratios for NGC\,5970.  Neither galaxy shows clear evidence for a systematic trend in spectral properties as a function of position angle.
}
\label{fig:mult_PA}
\end{figure}

Figure~\ref{fig:mult_PA} shows the negative results of this test, with no systematic variation in [N{\sc ii}] emission line EW being found with changing slit position angle.  For each angle, the point indicates the mean of the ratios observed on either side of the nucleus, and the ends of the error bars indicate the two individual values. The earlier-type NGC\,3185 has higher line ratios at all positions, but neither galaxy shows a significant difference in line ratios around the bar, that might indicate an `upstream' vs. `downstream' variation, and the values measured close to the bar do not show any convincing difference compared to those measured perpendicular to the bar.  Analysis of the [N{\sc ii}]/H$\alpha$ line ratio (not shown) also shows no evidence for variation with position relative to the bar. We conclude that there is no evidence for bar-driven shock excitation of this emission.  One consequence of this is that spectra taken with all slit orientations $\ge 15^{\circ}$ from the bar axis can be considered representative of the SFD, and will be included as such in statistical statements about SFD properties in this paper. However, given that only two galaxies were included in this analysis, no strong general statements can be made about the possible contribution of shocks to the diffuse line emission in SFD regions. The [O{\sc i}]6300\AA\ line, which gives stronger diagnostics of shock excitation, was not detected in any of the SFD regions. 

\subsection{Spectroscopic comparison of bar and SFD regions}

\begin{figure}
\includegraphics[width=90mm,angle=0]{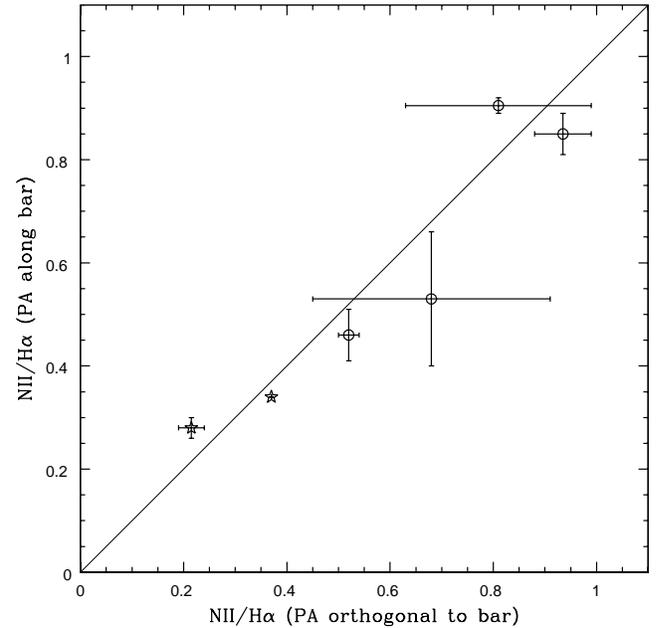}
\caption{
[N{\sc ii}]:H$\alpha$ line ratios for regions lying along the bar (y-axis)
vs. the same ratio extracted with the slit orientation orthogonal to the bar, within the same galaxies (x-axis), and covering the same radial ranges.
The H$\alpha$ fluxes have been corrected for the H$\alpha$ absorption 
as described in Section 3.2.}
\label{fig:bar_v_orthog}
\end{figure}

Figure~\ref{fig:mult_PA} shows a good agreement between the emission line strengths observed along and perpendicular to the bar for NGC\,3185 and NGC\,5970.  This raises the question of whether the spectroscopic properties of bars and SFD regions are generally similar.  Detailed analysis of this question will be left for a later paper, but we can attempt a preliminary analysis using spectroscopy for 6 galaxies in the present sample for which we have spectra observed along the bar axis, in addition to the SFD spectra.
Figure~\ref{fig:bar_v_orthog} shows the mean [N{\sc ii}]/H$\alpha$ ratios measured along the bars, plotted against the same ratios measured in the SFD regions perpendicular to the bar, for these 6 galaxies. The error bars indicate the ratios measured on opposite sides of the nucleus as previously, and the one-to-one relation is indicated by the diagonal line. In every case, the radial ranges of the spectral regions along and perpendicular to the bar are identical. 
The galaxies shown in this figure include two late-type, SBcd galaxies, NGC\,2604 and NGC\,4904, shown as stars in Fig.~\ref{fig:bar_v_orthog}.  The other four galaxies, plotted as circles are (reading from left to right on that figure) NGC\,5970 (type SBc), NGC\,5735 (SBbc), NGC\,3185 (SBa) and NGC\,2487 (SBc).

H{\sc ii} ratios, i.e. consistent with excitation by recently-formed stars, are found in the two Scd galaxies, NGC\,2604 and NGC\,4904, both along the bars and in the `SFD' regions.  Inspection of SDSS $gri$ colour images (not shown) confirms this, as both galaxies show very blue colours in their bars and the surrounding SFD regions.  The SFD regions show lower continuum surface brightness than the disc outside the bar-swept region, possibly indicating that the bars have started the removal of  material from their surroundings, but the blue colours and emission lines with SF ratios confirm that this has not yet suppressed SF. As a consequence, it does not make sense to talk about SF `deserts' in these galaxies, and this justifies the removal the 5 late-type Scd galaxies from the statistics discussed in Sect.\,3.3.

For the 8 bar regions remaining after removal of the Scd-types, again after the application of absorption corrections, 
the mean observed [N{\sc ii}]/H$\alpha$ ratio is 0.686$\pm$0.080, median
0.729.  A K-S test finds that these bar line ratios are completely consistent with being drawn from the same distribution as the ratios found in the SFD regions.  However, it is highly unlikely that they are drawn from the same distribution as the ratios measured from H{\sc ii} regions. 
The bar and SFD line ratio statistics confirm the visual impression from  Fig.~\ref{fig:bar_v_orthog}  that the bars in at least this sub-sample contain stellar population with similar spectroscopic properties to their surrounding disc regions.  However, this is a very small sample of galaxies and a very restricted set of properties, so we will return to this question in a later study.

\subsection{Sulphur line ratios as a diagnostic of electron density.}

\begin{figure}
\includegraphics[width=90mm,angle=0]{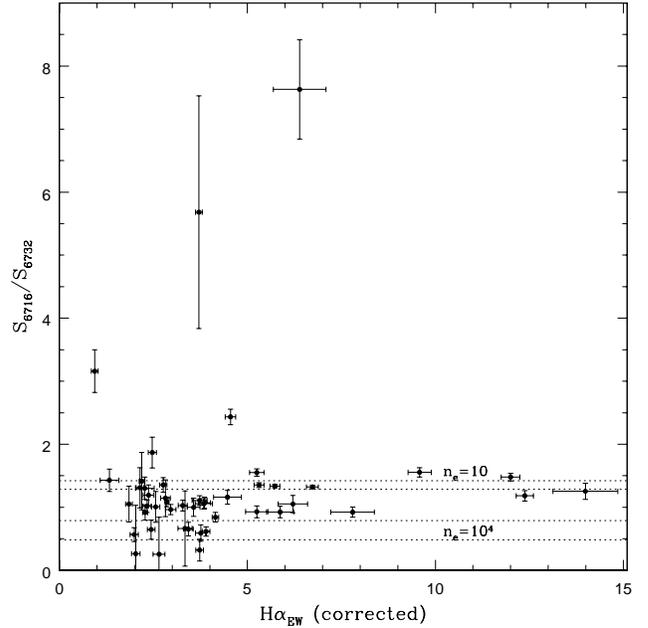}
\caption{The emission line ratio [S{\sc ii}]6716\AA/[S{\sc ii}]6732\AA\ as a function of H$\alpha$ equivalent width.  The dotted horizontal lines correspond to the predicted ratios for electron densities of 10, 100, 1000 and 10,000\,cm$^{-3}$ from top to bottom, respectively. 
}
\label{fig:SII_Ha}
\end{figure}

At the referee's suggestion, we have investigated whether we can derive useful constraints on ambient gas densities from our spectroscopic observations. \cite{prox14} present models of the emission line ratio [S{\sc ii}]6716\AA/[S{\sc ii}]6732\AA\ as a function of electron number density. These lines are detected at good signal-to-noise in most of our spectra.  In Fig.\,\ref{fig:SII_Ha}, we present the resulting estimates of electron number density as a function of H$\alpha$ EW.  The points with the highest EW values, above 7\,\AA, are predominantly late-type galaxies with SF-dominated line emission, as discussed previously (see Fig.\,\ref{fig:WHAN}). For the remaining population, dominated by LINER-type emssion, there is no evidence for a trend in line ratio, and hence $n_e$, with emission line strength.  Most of the regions lie within the range 100~$<n_e<$~1000, with outliers both above and below this range.

\section{Discussion}

Several results emerge from this analysis of 48 barred galaxies.  We confirm our earlier findings that strong bars have a profound effect on star formation in the central few kpc of galaxies. With the exception of the five SBcd galaxies, all of the galaxies studied show the SFD phenomenon, with no evidence of SF-driven emission line ratios in the regions swept by the bars of these 43 spiral galaxies.  However, we also confirm the presence of diffuse line emission from almost all of these SFD regions, as first reported in \cite{jame15}.  The smooth morphology of this emission is very different from the clumpy emission that characterises H{\sc ii} regions, and favours an association with the older stellar population that is present throughout these SFD regions. It extends several kpc from the galaxy centres, arguing against AGN as the powering mechanism.

This diffuse emission is close to ubiquitous, with [N{\sc ii}] line emission observed in 93 out of 101 SFD regions observed (92 per cent of SFD region spectra).  The fraction with H$\alpha$ line emission is at least as high with 96 regions showing emission, after correction for underlying H$\alpha$ absorption. The distributions of [N{\sc ii}]/H$\alpha$ line ratio (Fig.~\ref{fig:ratios_hist} lower frame), H$\alpha$ EW (Fig.~\ref{fig:WHAN}) and [N{\sc ii}] EW all show quite narrow distributions, with no obvious bimodality.  This ubiquity and uniformity of spectroscopic properties favours p-AGB over EHB stars as the ionising source. While there are no direct constraints on the presence or absence of EHB stars in the SFD regions, evidence from resolved populations in globular clusters \citep{cate09,tore19} shows that only about one-half of these systems have horizontal branches that extend to very blue colours.  In a galactic context, the UV-upturn phenomenon, which is believed to be driven by EHB stars \citep{cart11}, is found to vary greatly in strength from galaxy to galaxy, and is strongest in the centres of bright elliptical galaxies with high [$\alpha$/Fe] abundance ratios, a very different environment to the SFD regions of spiral galaxies.  The uniformity of the line emission properties found here is much more easily explained if the emission is associated with p-AGB stars, which are themselves likely to be ubiquitous in intermediate-age and old stellar populations.  Hence we take this as our working assumption for the rest of this discussion.

When we compare our observations of SFD regions with the p-AGB emission line models of \cite{byle17,byle19}, some striking agreements are found, but there are also some interesting discrepancies.  The [N{\sc ii}]/H$\alpha$ line ratio distribution shown in the lower frame of Fig.~\ref{fig:ratios_hist} shows quite a tight clustering of values around a mean of 0.648.  This is in excellent agreement with the `Fiducial' model shown in fig.~8 of \cite{byle19}, which predicts a logarithmic [N{\sc ii}]/H$\alpha$ ratio of --0.2, or a ratio of 0.63, where the gas excited by the p-AGB stars has solar metallicity and abundance ratios.  \cite{byle19} introduce a modified set of models, with enhanced abundances of $\alpha$ elements, carbon and nitrogen (termed $\alpha$CN models) to simulate the effects of the local gas being dominated by gas ejected by the earlier evolution of the AGB stars themselves.  They find that these $\alpha$CN models successfully reproduce the line ratios characteristic of the LINER/LIER region of the \cite{bald81} and \cite{kewl06} diagnostic diagrams, including [N{\sc ii}]/H$\alpha$ ratios of approximately 2.0 -- 3.0.  However, our SFD spectra are clearly inconsistent with these $\alpha$CN model predictions, so we conclude that the observed [N{\sc ii}]/H$\alpha$ line ratios are best explained by p-AGB stars ionising ambient interstellar medium, which must have sufficient density to dominate over AGB wind material.  The presence of significant amounts of ambient gas may also explain why the line emission is marginally stronger (2 -- 4\,\AA\ EW in H$\alpha$, see Fig.~\ref{fig:WHAN}) compared with the model predictions (0.2 - 2.5\,\AA). Unfortunately, our analysis of S{\sc ii} emission line ratios does not show any clear evidence for a dependence of the strength of this LINER-type emission on local gas density as traced by $n_e$.  Further work on this question, particularly a comparison with stellar populations in passive early-type galaxies, would be instructive.

A controlling role for the gas supply in related contexts has been suggested previously.
\cite{grav07}, in an analysis of nuclear SDSS spectroscopy (and thus rather different environments from the present study), find LINER-type emission to be common but not ubiquitous in early-type galaxies.  They conclude that significant ionising flux should be provided by p-AGB stars in all old stellar populations, and that this is the likely source powering such extended LINER emission; those galaxies without such emission are then explained through the absence of gas.  Comparing with the WHAN diagnostic diagrams presented by\cite{cidf11}, who again analyse nuclear SDSS spectra to discriminate emission powered by star formation, AGN and hot evolved stars, we find that our SFD regions lie in a region that is clearly in the LINER regime, but that is relatively poorly populated by nuclear regions.  Thus the SFD regions lie below the EW values expected for star formation or strong AGN, but higher than those found by \cite{cidf11} for red sequence `retired' and passive galaxies.  Again, this indicates that SFD regions have emission properties quite unlike those of least the central regions of any other galaxies, due to a combination of suppressed star formation, but higher ambient gas densities than are found in most quenched galaxies.

However, there is clear evidence, both from observations and simulations, that bars do deplete gas in  SFD regions. In a study of simulations of strongly-barred spiral galaxies, \cite{dono19} show significant removal of gas from SFD regions, suppressing star formation, within $\sim$1~Gyr of the formation of the bar.  This finding is confirmed, for one galaxy at least, by the observational study of \cite{geor19}, who present an analysis of the gas properties of the very nearby SBb galaxy NGC\,3351, which was included in \cite{jame15} and hence also in the present study.  \cite{geor19} found that the SFD region, and most of the bar itself, are significantly depleted in both atomic and molecular gas, as traced by 21\,cm radio and CO line emission respectively. Thus the bar appears to be funnelling gas from the entire region surrounding the bar into the nucleus of NGC\,3351.  However, in the present paper, we do find [N{\sc ii}] emission from the SFD region of this galaxy, and H$\alpha$ emission after application of the H$\alpha$ absorption correction.  Thus there must be some gas present, as this line emission is clearly stronger than that from heavily gas-depleted passive galaxies \citep[e.g.][]{cidf11}.  In this context, it is interesting to note that the simulation-based study of \cite{dono19} finds that, even after bars and SFD regions are fully established, significant radial migration occurs, bringing both stars and gas from the outer disc into the SFD regions.  In the simulations, this radial flow does not re-establish star formation in the SFD regions, but it might be sufficient to explain the emission line properties found here. 

DIG, excited by ionizing photons leaking from H{\sc ii} regions is likely to make at least some  contribution to the diffuse extended emission discussed here.  However, we feel that it is not the most natural explanation for the majority of this emssion, over the full extent of these SFD regions.  While \cite{zuri02} describe the DIG as and extended component, their models show it as still be generally concentrated around star-forming complexes as an extended halo (see e.g. their Fig.~4).  In the SFD regions of the present study, the emission is smoothly distributed over  the full extent of the regions, covering many kpc, with no evidence for any H{\sc ii} regions within the SFD regions.  In addition, while \cite{belf16} find evidence in their IFU observations for LINER-like ratios from DIG emission, they also find that such emission is strongest close to regions of SF, and that the line ratios cluster on the borderline between LINER-like and SF excitation.  \cite{zhan17} take this further, concluding, that `Leaky H{\sc ii} region models cannot produce LI(N)ER- like emission...LI(N)ER-like emission needs another ionization source'. Thus we focus in the present work on pAGB excitation as our preferred interpretation for the dominant excitation source in the regions observed in this paper.  However, there is an interesting debate to be had on the size and fraction of the contribution made by DIG, and we emphasise the importance of further observations of these SFD regions, where the extended scales may give unique insights into the DIG and pAGB contributions.

In summary, we find an important role for gas supply as a controlling factor in the observed properties of these SFD regions.  While they are expected to be sufficiently gas-depleted so as to completely suppress star formation, the strength and line ratios of the diffuse line emission is most easily explained by p-AGB stars surrounded by significant amounts of ambient gas of approximately solar abundance. We note that there is support for this picture in the unusual properties of a post-merger barred galaxy, NGC\,3729, described in Appendix A.  Further study is required to determine why the gas in these SFD regions does not result in star formation.

\section{Conclusions}

The main conclusions of the present study are as follows:

\begin{itemize}
\item We confirm the  SFD phenomenon of local suppression of star formation by bars, in the largest sample to date, 48 galaxies.
\item The phenomenon occurs in virtually all early-type spirals (SBa - SBc), but 5 SBcd galaxies in the current sample retain star formation throughout their bars and surrounding regions.
\item Diffuse H$\alpha$ and [N{\sc ii}] line emission is seen in almost all SFD regions, with [N{\sc ii}]/H$\alpha$ line ratios of 0.65 being typical, and EWs of 1.5 -- 4\,\AA.
\item The ubiquity and homogeneity of the line emission across all of the galaxies studied here argues for p-AGB stars, rather than e.g. EHB stars, as the dominant ionisation source.
\item Line ratios are consistent with models of excitation of solar-abundance gas by p-AGB stars.
\item The strength of the emission lines argues for a higher ambient gas density than is found in passive red-sequence galaxies.
\item A preliminary comparison of the emission-line properties of bars and SFD regions finds little difference between the two environments within a given galaxy.
\item Further study is needed to determine why the SFD region gas content identified here results in no detectable star formation.
\end{itemize}

In many ways, the SFD regions seem to represent a unique galactic environment, both in terms of their star formation properties, and in the emission lines resulting from their evolved stellar populations. 
 
\section*{Acknowledgments}

We thank the anaonymous referee for several suggestions which significantly improved the paper.  The Isaac Newton Telescope is operated on the island of La Palma by the Isaac Newton Group in the Spanish Observatorio del Roque de los Muchachos of the Instituto de Astrof\'isica de Canarias.
This research has made use of the NASA/IPAC Extragalactic Database (NED) which is operated by the Jet Propulsion Laboratory, California Institute of Technology, under contract with the National Aeronautics and Space Administration.  
This research has made use of the
NASA/IPAC Extragalactic Database (NED) which is operated by the Jet Propulsion Laboratory, California Institute
of Technology, under contract with the National Aeronautics and Space Administration. Funding for the SDSS and
SDSS-II has been provided by the Alfred P. Sloan Foundation, the Participating Institutions, the National Science
Foundation, the U.S. Department of Energy, the National
Aeronautics and Space Administration, the Japanese Monbukagakusho, the Max Planck Society, and the Higher Education Funding Council for England. The SDSS Web Site
is http://www.sdss.org/. The SDSS is managed by the Astrophysical Research Consortium for the Participating Institutions. The Participating Institutions are the American
Museum of Natural History, Astrophysical Institute Potsdam, University of Basel, University of Cambridge, Case
Western Reserve University, University of Chicago, Drexel
University, Fermilab, the Institute for Advanced Study, the
Japan Participation Group, Johns Hopkins University, the
Joint Institute for Nuclear Astrophysics, the Kavli Institute for Particle Astrophysics and Cosmology, the Korean
Scientist Group, the Chinese Academy of Sciences (LAMOST), Los Alamos National Laboratory, the Max-PlanckInstitute for Astronomy (MPIA), the Max-Planck-Institute
for Astrophysics (MPA), New Mexico State University, Ohio
State University, University of Pittsburgh, University of
Portsmouth, Princeton University, the United States Naval
Observatory, and the University of Washington.


\begin{thebibliography}{}

\bibitem[\protect\citeauthoryear{Allen et al.}{2008}]{alle08} Allen M.~G., Groves B.~A., Dopita M.~A., Sutherland R.~S., Kewley L.~J., 2008, ApJS, 178, 20

\bibitem[\protect\citeauthoryear{Baldwin, Phillips, 
\& Terlevich}{1981}]{bald81} Baldwin J.~A., Phillips M.~M., Terlevich R., 1981, PASP, 93, 5 

\bibitem[\protect\citeauthoryear{Belfiore et al.}{2016}]{belf16} Belfiore F.et al., 2016, MNRAS, 461, 3111 

\bibitem[\protect\citeauthoryear{Binette et al.}{1994}]{bine94} Binette L., Magris C.~G., Stasi{\'n}ska G., Bruzual A.~G., 1994, A\&A, 292, 13 

\bibitem[\protect\citeauthoryear{Bremer et 
al.}{2013}]{brem13} Bremer M., Scharw{\"a}chter J., Eckart A., Valencia-S.~M., Zuther J., Combes F., Garcia-Burillo S., Fischer S., 2013, A\&A, 558, A34 

\bibitem[\protect\citeauthoryear{Brocklehurst}{1971}]{broc71} Brocklehurst M., 1971, MNRAS, 153, 471 

\bibitem[\protect\citeauthoryear{Byler et al.}{2017}]{byle17} Byler N., Dalcanton J.~J., Conroy C., Johnson B.~D., 2017, ApJ, 840, 44 

\bibitem[\protect\citeauthoryear{Byler et al.}{2019}]{byle19} Byler N., Dalcanton J.~J., Conroy C., Johnson B.~D., Choi J., Dotter A., Rosenfield P., 2019, arXiv, arXiv:1904.10978

\bibitem[\protect\citeauthoryear{Caldwell}{1984}]{cald84} Caldwell N., 1984, PASP, 96, 287 

\bibitem[\protect\citeauthoryear{Carter, Pass, Kennedy, Karick \& Smith}{2011}]{cart11} Carter D., Pass S., Kennedy J., Karick A.~M., Smith R.~J., 2011, MNRAS, 414, 3410

\bibitem[\protect\citeauthoryear{Catelan}{2009}]{cate09} Catelan M., 2009, Ap\&SS, 320, 261

\bibitem[\protect\citeauthoryear{Choi et al.}{2016}]{choi16} Choi J., Dotter A., Conroy C., Cantiello M., Paxton B., Johnson B.~D., 2016, ApJ, 823, 102 

\bibitem[\protect\citeauthoryear{Cid Fernandes et al.}{2011}]{cidf11} Cid Fernandes R., Stasi{\'n}ska G., Mateus A., Vale Asari N., 2011, MNRAS, 413, 1687 

\bibitem[\protect\citeauthoryear{den Brok et al.}{2020}]{denb20} den Brok M., et al., 2020, MNRAS, 491, 4089

\bibitem[\protect\citeauthoryear{Donohoe-Keyes et al.}{2019}]{dono19} Donohoe-Keyes C.~E., Martig M., James P.~A., Kraljic K., 2019, MNRAS, 489, 4992

\bibitem[\protect\citeauthoryear{Dotter}{2016}]{dott16} Dotter A., 2016, ApJS, 222, 8 

\bibitem[\protect\citeauthoryear{Ferguson et al.}{1996}]{ferg96} Ferguson A.~M.~N., Wyse R.~F.~G., Gallagher J.~S., Hunter D.~A., 1996, AJ, 111, 2265

\bibitem[\protect\citeauthoryear{Ferland et al.}{2013}]{ferl13} Ferland G.~J., et al., 2013, RMxAA, 49, 137 

\bibitem[\protect\citeauthoryear{Friedli \& Benz}{1993}]{frie93} Friedli D., Benz W., 1993, A\&A, 268, 65

\bibitem[\protect\citeauthoryear{George et al.}{2019}]{geor19} George K., Joseph P., Mondal C., Subramanian S., Subramaniam A., Paul K.~T., 2019, A\&A, 621, L4 

\bibitem[\protect\citeauthoryear{Gomes et al.}{2016}]{gome16} Gomes J.~M., et al., 2016, A\&A, 588, A68 

\bibitem[\protect\citeauthoryear{Goudfrooij}{1999}]{goud99} Goudfrooij P., 1999, in Carral P., Cepa J., eds, Star Formation in Early-type Galaxies, ASP Conf. Ser. Vol. 163. Astronomical Society of the Pacific, p. 55 

\bibitem[\protect\citeauthoryear{Graves et al.}{2007}]{grav07} Graves G.~J., Faber S.~M., Schiavon R.~P., Yan R., 2007, ApJ, 671, 243 

\bibitem[\protect\citeauthoryear{Haffner et al.}{2009}]{haff09} Haffner L.~M., et al., 2009, RvMP, 81, 969

\bibitem[\protect\citeauthoryear{Hakobyan et 
al.}{2014}]{hako14} Hakobyan A.~A., et al., 2014, MNRAS, 444, 
2428 

\bibitem[\protect\citeauthoryear{Hirschmann et al.}{2017}]{hirs17} Hirschmann M., Charlot S., Feltre A., Naab T., Choi E., Ostriker J.~P., Somerville R.~S., 2017, MNRAS, 472, 2468 

\bibitem[\protect\citeauthoryear{Hummer \& Storey}{1987}]{humm87} Hummer D.~G., Storey P.~J., 1987, MNRAS, 224, 801 

\bibitem[\protect\citeauthoryear{James et al.}{2004}]{jame04} 
James, P.~A. et al., 2004, A\&A, 414, 23

\bibitem[\protect\citeauthoryear{James et al.}{2009}]{jame09} 
James P.~A., Bretherton C.~F. \& Knapen J.~H., 2009, A\&A, 501, 207

\bibitem[\protect\citeauthoryear{James \& Percival}{2015}]{jame15} James P.~A., Percival S.~M., 2015, MNRAS, 450, 3503 

\bibitem[\protect\citeauthoryear{James \& Percival}{2016}]{jame16} James P.~A., Percival S.~M., 2016, MNRAS, 457, 917 

\bibitem[\protect\citeauthoryear{James \& Percival}{2018}]{jame18} James P.~A., Percival S.~M., 2018, MNRAS, 474, 3101 

\bibitem[\protect\citeauthoryear{Kewley, Groves, Kauffmann \& Heckman}{2006}]{kewl06} Kewley L.~J., Groves B., Kauffmann G., Heckman T., 2006, MNRAS, 372, 961

\bibitem[\protect\citeauthoryear{Kormendy et al.}{2004}]{korm04} 
Kormendy J., \& Kennicutt R.~C., 2004, ARAA, 42, 603

\bibitem[\protect\citeauthoryear{Kormendy}{2013}]{korm13} Kormendy J., 2013, in Falcon-Barroso J., Knapen J. H., eds, Secular Evolution of Galaxies, XXIII Canary Islands Winter School of Astrophysics, Cambridge University Press, Cambridge, p. 1

\bibitem[\protect\citeauthoryear{Nair 
\& Abraham}{2010}]{nair10} Nair P.~B., Abraham R.~G., 2010, ApJ, 714, L260

\bibitem[\protect\citeauthoryear{Papaderos et al.}{2013}]{papa13} Papaderos P., et al., 2013, A\&A, 555, L1

\bibitem[\protect\citeauthoryear{Percival et al.}{2009}]{perc09} 
Percival S.~M., Salaris M., Cassisi S. \& Pietrinferni A.,2009, ApJ, 690, 427 

\bibitem[\protect\citeauthoryear{Phillips et al.}{1986}]{phil86} Phillips M.~M., Jenkins C.~R., Dopita M.~A., Sadler E.~M., Binette L., 1986, AJ, 91, 1062 

\bibitem[\protect\citeauthoryear{Proxauf, {\"O}ttl \& Kimeswenger}{2014}]{prox14} Proxauf B., {\"O}ttl S., Kimeswenger S., 2014, A\&A, 561, A10

\bibitem[\protect\citeauthoryear{Singh et al.}{2013}]{sing13} Singh R., et al., 2013, A\&A, 558, AA43 

\bibitem[\protect\citeauthoryear{Stasi{\'n}ska et 
al.}{2008}]{stas08} Stasi{\'n}ska G., et al., 2008, MNRAS, 
391, L29 

\bibitem[\protect\citeauthoryear{Torelli et al.}{2019}]{tore19} Torelli M., et al., 2019, arXiv, arXiv:1907.09568

\bibitem[\protect\citeauthoryear{Weilbacher et al.}{2018}]{weil18} Weilbacher P.~M., et al., 2018, A\&A, 611, A95

\bibitem[\protect\citeauthoryear{Zhang et al.}{2017}]{zhan17} Zhang K., et al., 2017, MNRAS, 466, 3217

\bibitem[\protect\citeauthoryear{Zurita et al.}{2002}]{zuri02} Zurita A., Beckman J.~E., Rozas M., Ryder S., 2002, A\&A, 386, 801

\end{thebibliography}


\appendix

\section{An unusual merging barred galaxy NGC\,3729}

\begin{figure}
\includegraphics[width=85mm,angle=0]{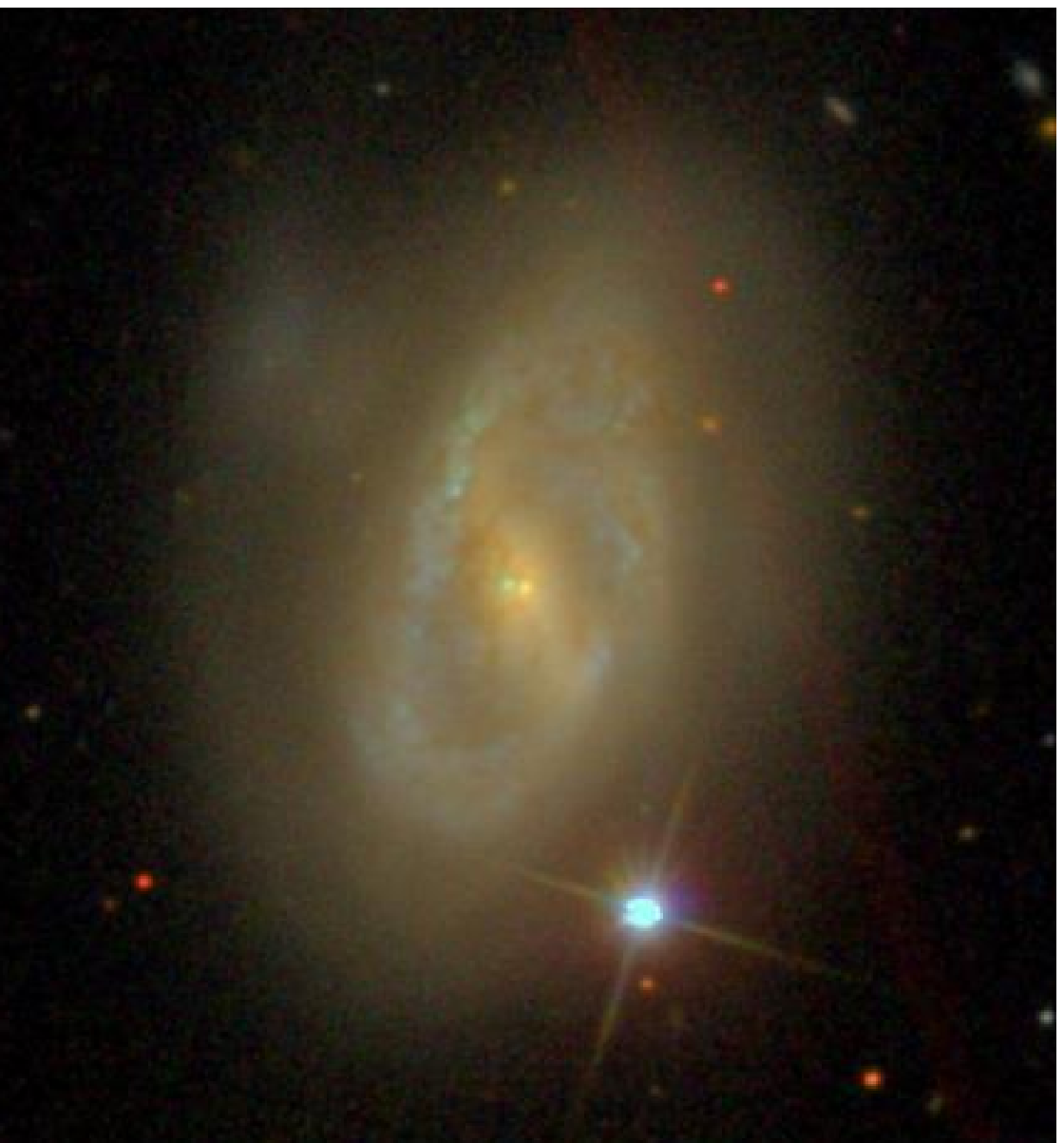}
\caption{SDSS colour composite image on NGC\,3729.  Note the low surface brightness component to the north-east, which may indicate an ongoing minor merger. 
}
\label{fig:ngc3729_sdss}
\end{figure}

\begin{figure}
\includegraphics[width=90mm,angle=0]{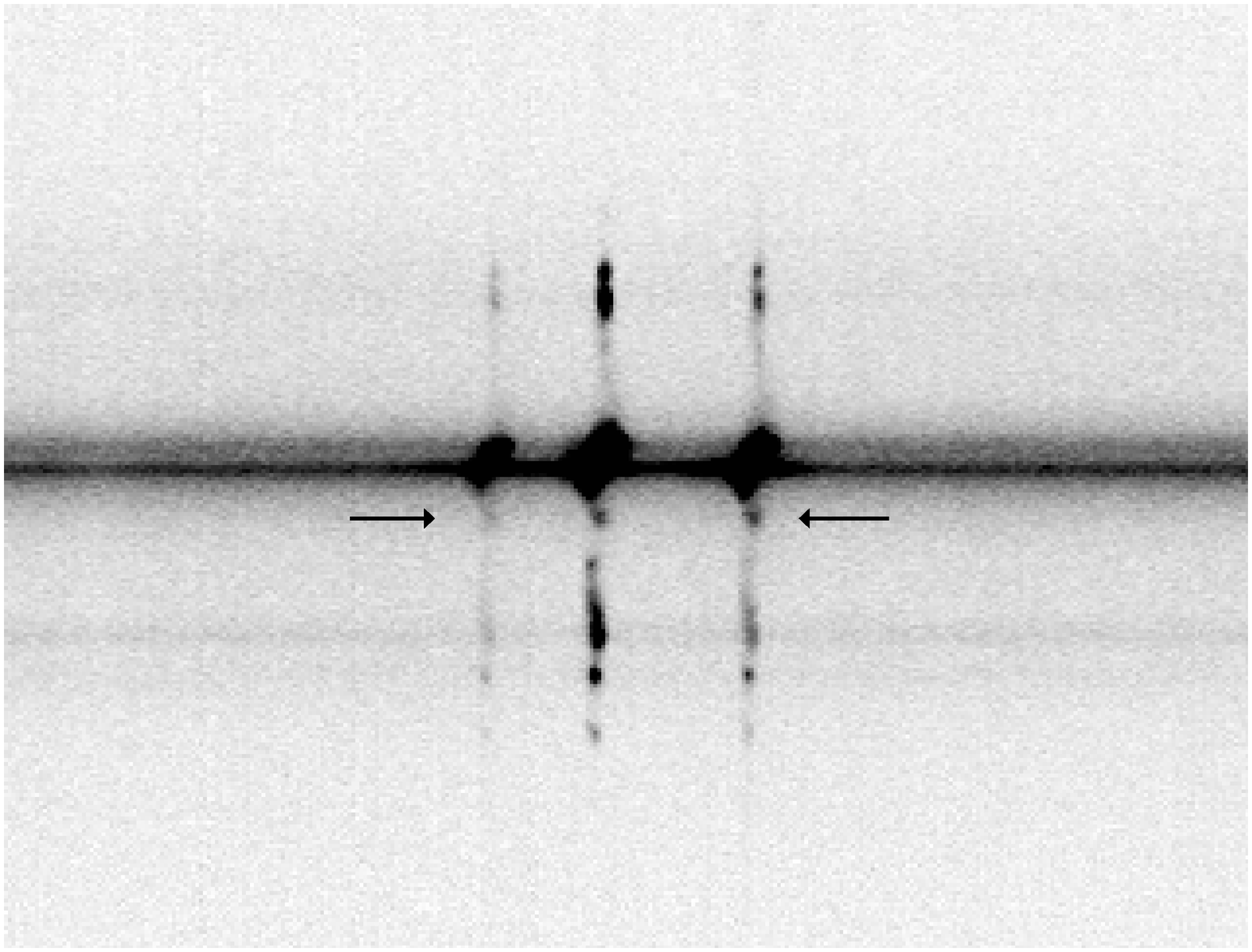}
\caption{Two-dimensional optical spectrum of NGC\,3729, centred on the H$\alpha$ and [N{\sc ii}] lines. The emission-line clump is indicated by the arrows.  Comparison with the bright H{\sc ii} region towards the bottom of the frame shows the extended nature of the line emission from the clump. 
}
\label{fig:ngc3729spec}
\end{figure}

\begin{figure}
\includegraphics[width=90mm,angle=0]{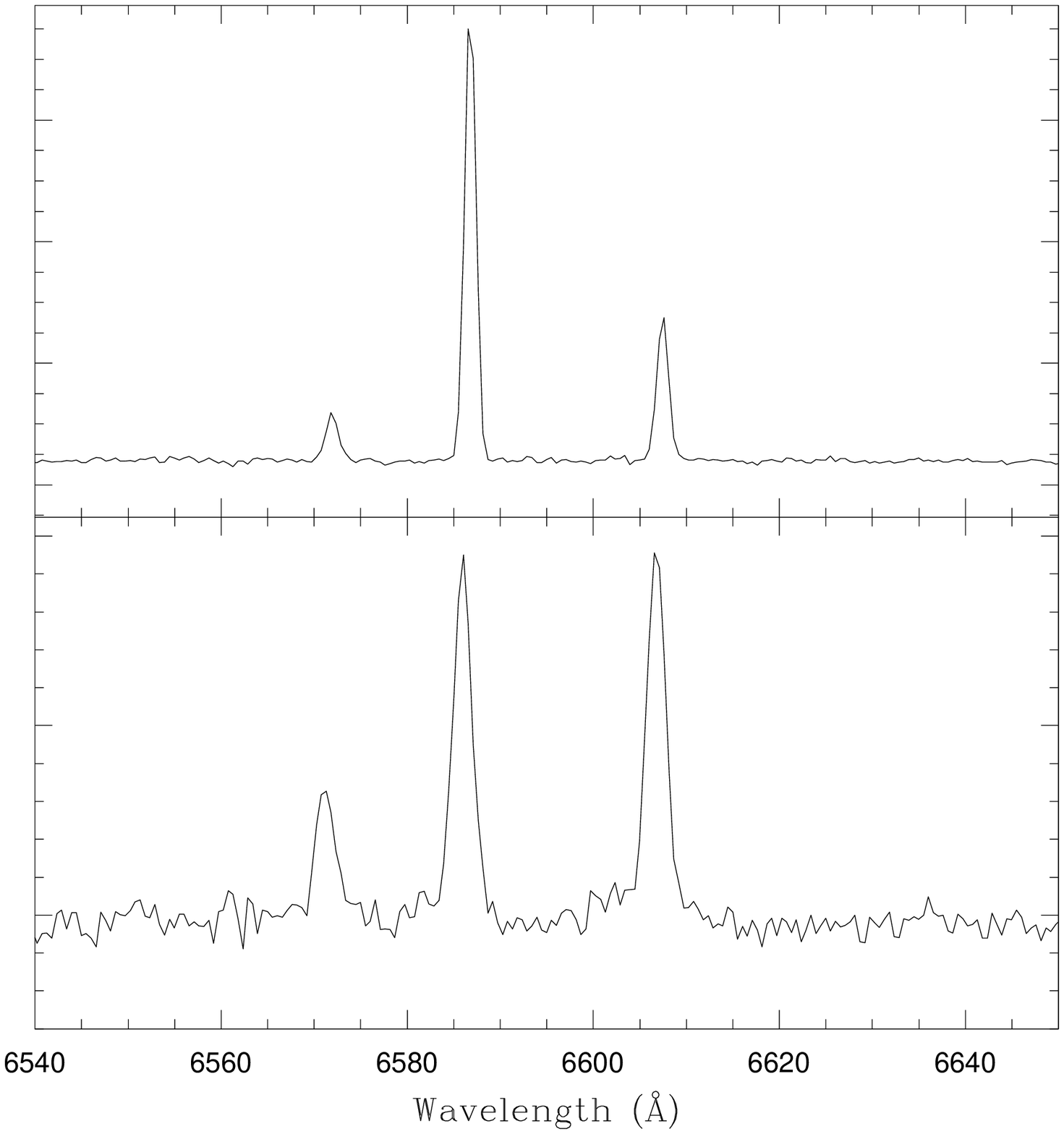}
\caption{Extracted H$\alpha$ and [N{\sc ii}] emission for the NGC\,3729 clump (bottom) and, for comparison, an H{\sc ii} region in the same galaxy (top). The high [N{\sc ii}]/H$\alpha$ ratio in the clump is clearly apparent. 
}
\label{fig:ngc3729_2spec}
\end{figure}

In this Appendix, we describe a somewhat unusual barred galaxy, NGC~3729. This has a classification of SBa (pec), with the peculiar nature being linked to a probable late-stage minor merger, as illustrated in the SDSS colour composite image in Fig.~\ref{fig:ngc3729_sdss}. The low-surface-brightness feature to the north-east of NGC~3729 may be the tidally-shredded remains of a low-mass companion, and the mottled appearance of the upper half of NGC~3729 may be due to dusty gas accreted during this interaction.  We adopt a distance of 20.2~Mpc for this system.

Spectroscopically, NGC~3729 shows strong and complex central emission, and emission from a bar-end ring, all of which has [N{\sc ii}]/H$\alpha$ ratios fully consistent with star formation; see the 2-dimensional spectrum in Fig.~\ref{fig:ngc3729spec}, taken with the INT and IDS at a slit position angle of 290$^{\circ}$, perpendicular to the bar.  The bar-end ring results in the strong peaks of emission towards the top and bottom of this figure, at distance of 19$^{\prime\prime}$ or 1.8~kpc from the nucleus.
The spectrum shows complex nuclear structure, with multiple components, including a very rapidly rotating structure, which we interpret as a star-forming nuclear ring.  Between the nuclear and bar-end rings, diffuse emission is apparent, particularly in the upper part of Fig.~\ref{fig:ngc3729spec}.  Thus, in spite of its peculiarity, we do consider this an example of the SFD phenomen, and it is included as such in the present paper.

The unusual feature seen in Fig.~\ref{fig:ngc3729spec} is the peak of emission seen just below the nuclear ring, 5.7$^{\prime\prime}$ (560~pc) from the nucleus.  At first sight, this resembles an H{\sc ii} region, but uniquely, in the current data set at least, it shows clear LINER-type ratios.  To illustrate this point, Fig.~\ref{fig:ngc3729_2spec} shows the extracted spectrum of this peak in the lower frame, with an H{\sc ii} region spectrum from the same long-slit observation shown in the upper frame for comparison. The spectra are clearly different, and argue against ongoing star formation in the near-nuclear clump.  The clump has [N{\sc ii}] and H$\alpha$ EW values of 5.63 and 6.73\,\AA\ respectively, similar to or slightly greater than the values for the two SFD regions in NGC\,3729 (4.70 and 3.69\,\AA\ for [N{\sc ii}], 6.76 and 4.50\,\AA\ for H$\alpha$). These values are all at the high end of the EW distributions for line emission from SFD regions, while the [N{\sc ii}]/H$\alpha$ ratio for the clump, 0.837, is entirely consistent with the distribution of ratios for SFD regions (Fig.~\ref{fig:ratios_hist} lower panel). 

One possibility is that the near-nuclear clump could be the nucleus of the tidally-shredded companion, that has been drawn towards the centre of NGC~3729 through dynamical friction.  One test of this is to look for spatial extent. Visual inspection of Fig.~\ref{fig:ngc3729spec} indicates that the clump  appears moderateley extended, e.g. compared with the most compact H{\sc ii} region (towards the bottom of the image).  Measurements confirm this; along the spatial direction, the clump has a full-width at half maximum (FWHM) of 5.3 pixels, or 2.33$^{\prime\prime}$, compared to 3.7 pixels or 1.63$^{\prime\prime}$ for the compact H{\sc ii} region.  As the latter may itself be somewhat extended, we also measured the FWHM of the standard star observation taken closest in time to the NGC\,3729 spectra, finding values of 3.2 pixels or 1.4$^{\prime\prime}$.  Taking either the star or (conservatively) the H{\sc ii} region as a measure of the seeing broadening the clump FWHM, the true size of the clump is 1.9 or 1.7$^{\prime\prime}$ respectively. Taking the distance to the galaxy as 20.2\,Mpc, the FWHM size of the clump is 163 - 182\,pc, much larger than a stellar cluster.  Thus this may be the extended central regions of a dwarf galaxy.
We also note that the clump velocity is somewhat different from that of the disc and nuclear emission to either side of it, as might be expected if the clump were part of an interacting or merging system.

We note that NGC\,3729 may well be an unusually gas-rich galaxy within the current sample, due to the merging activity, and evidenced by the clumpy absorption seen over much of the galaxy.  This may explain the stronger-than-average line emission seen in the SFD regions of this galaxy. The gas would be expected to concentrate in the central regions of the galaxy, explaining the particularly high EW lines from the near-nuclear clump. Overall, the spectroscopic properties of the clump in NGC\,3729 support the interpretation of diffuse line emission presented in this paper, with a ubiquitous source of ionising photons from old stellar populations in all locations, and the EW of the resulting line emission being controlled primarily by the gas supply.  The line ratios in the clump are consistent with theoretical predictions where ambient gas of approximately solar abundance is dominant, as was found for the SFD regions in the present study. 


\label{lastpage}

\end{document}